\documentclass{emulateapj}
\usepackage{epsf}
\usepackage{natbib}
\usepackage{apjfonts}
\usepackage{epsfig}

\newcommand{\alphaFet}{[$\alpha$/Fe]}
\newcommand{\alphaFe}{[$\alpha$/Fe]\,}
\newcommand{\FeHt}{[Fe/H]}
\newcommand{\FeH}{[Fe/H]\,}
\newcommand{\avFeH}{$<$[Fe/H]$>$\,}

\shorttitle{Phase-Space Distributions of Chemical Abundances}
\shortauthors{Font et~al.}

\begin{document}

\title{Phase-Space Distributions  of Chemical Abundances in Milky Way-type Galaxy Halos}

\author{
Andreea~S.~Font\altaffilmark{1,*},
Kathryn~V.~Johnston\altaffilmark{1},
James~S.~Bullock\altaffilmark{2},
Brant~E.~Robertson\altaffilmark{3}
}

\altaffiltext{1}{Van Vleck Observatory, Wesleyan University, Middletown, 
CT 06459, USA}
\altaffiltext{2}{Center for Cosmology, Department of Physics \& Astronomy, University of California, 
Irvine, CA 92687, USA}
\altaffiltext{3}{Harvard-Smithsonian Center for Astrophysics, 60 Garden Street, 
Cambridge, MA 02138, USA}
\altaffiltext{*}{afont@astro.wesleyan.edu}

\begin{abstract}

Motivated by upcoming data  from astrometric and spectroscopic surveys
of  the  Galaxy, we   explore  the  chemical abundance  properties and
phase-space distributions   in  hierarchically-formed   stellar   halo
simulations set  in a $\Lambda$CDM Universe.   Our sample of Milky-Way
type stellar  halo  simulations result  in average metallicities
that range  from  \FeH $\simeq -1.3$ to $-0.9$,  with  the most metal  poor  halos
resulting from  accretion   histories  that lack destructive mergers with
massive   (metal    rich) satellites.   Our stellar halo
metallicities increase with stellar halo mass. The slope of the \FeH $-
M_{*}$ trend  mimics   that  of  the  satellite  galaxies that were
destroyed to build the halos, implying that  the relation propagates  hierarchically.  All simulated
halos  contain a  significant   fraction of  old  stellar  populations
accreted more  than 10~Gyr ago and  in a  few cases, some intermediate
age populations exist.   In contrast with  the Milky Way,  many of our
simulated  stellar  halos  contain old stellar   populations which are
metal  rich, originating in  the early accretion of massive satellites
($M_{*}  \sim 10^{9}  M_{\odot}$).  We  suggest  that  the  (metal rich)
stellar halo of  M31 falls into this  category,  while the  more metal
poor halo  of  the Milky Way  is  lacking in   early massive accretion
events.  Interestingly, our  hierarchically-formed stellar halos often
have  non-negligible   metallicity   gradients  in    both  \FeH   and
\alphaFet. These gradients extend a few tens of kpc, and can be as large as
0.5~dex in \FeH and 0.2~dex in \alphaFet, with the most metal poor halo
stars typically buried within the central  $\sim 5$~kpc of the galaxy.
Thus evidence for metallicity gradients alone in the  Milky Way stellar halo
would not preclude   its formation via a   hierarchical process.  Only
coupled with phase-space data  can metallicity information be utilized
to test ideas about accreted versus in situ formation.  
Finally, we  find that chemical abundances can
act as a rough substitute for  time of accretion of satellite galaxies
and, based on   this finding, we  propose a  criterion for identifying
tidal streams spatially by  selecting stars with \alphaFe ratios below
solar.

\end{abstract}


\section{Introduction}
\label{sec:introduction}

The  stars in the halo  hold important information about the formation
history of the Galaxy. Their spatial and velocity distributions can be
used  to retrace their dynamic origin, and their
chemical   abundances can  constrain   the  star  formation histories
of their constituents.
A major goal over the next decade is to obtain kinematic and chemical
information for a large number of stars in our own Galaxy, thus
providing a detailed reconstruction of its formation history and a link
to the underlying cosmology \citep[e.g.][]{freeman02}.  It is the goal
of this work to provide a fist step towards the modeling needed to exploit
these observations to their fullest potential.

The combination of spatial, kinematic and chemical data will provide a powerful
discriminate   for   testing  different   Galaxy  formation    models.
Historically,  metallicity 
gradients have acted a primary motivator in formulating
hypotheses:   large-scale
metallicity gradients are associated with a rapid, semi-continuous collapse
\citep[][ELS]{eggen62}; homogeneous metallicity distributions
are associated with  chaotic assembly from fragments
\citep[][SZ]{searle77,searle78}.  While there are no well-motivated 
models that predict these opposite extremes from first principles,
the poles of debate set by ELS and SZ serve as important straw-man
theories for shaping ideas and testing data.  
Nevertheless, realistic models will always fall into less idealized categories.
For example, our $\Lambda$CDM-based model relies entirely on a hierarchical
origin for the stellar halo, yet predicts some non-negligible
metallicity gradients (see below).  Likewise
models that allow significant
``rapid'' in situ star formation necessarily contain some accreted
stellar material as a result of having a background hierarchical cosmology
\cite[e.g.][]{renda05}.  Clearly, only in combination with
spatial and  kinematic data can chemical abundance data be used
to its full potential.
 
Numerous  results over the  past decades support  the  validity of the
hierarchical   model of  structure   formation \citep{white78,blumenthal84}.  
In  this  model,  the expectation for stellar halos of
galaxies like the Milky Way is that they form
in large part by   phase-mixing of    tidal debris   from  the numerous     accreted
satellites. Although the dynamical  evolution of galaxy halos has been
extensively modeled, and we have  now predictions for the  phase-space
distribution of both dark matter and  stars in the  halo of the Galaxy
\citep[e.g.][]{johnston98,helmi99,bullock01,helmi03,abadi05,bullock05,diemand05,moore05},
currently    there are  no    clear   predictions for  the  associated
phase-space distribution of chemical elements.

Intuitively, one expects   stochastic  accretions  to leave   peculiar
signatures in both the  kinematics and chemical abundances of  present
day stars. Some observational evidence in that respect is indeed found
in  our  Galaxy.  The halo contains  stars  which  stand out  in  both
metallicity                       and                         velocity
\citep{carney96,nissen91,nissen97,majewski96,chiba00,altmann05},    as
does the  disk \citep{helmi99,navarro04,helmi05}. Similar evidence has
been found in M31 $-$ the only  other spiral galaxy whose stellar halo
has been studied at a comparable level of detail  as that of the Milky
Way
\citep{ferguson02,reitzel02,bellazzini03,ferguson05,guhathakurta05a}. The
giant stellar stream   in M31, a merger   debris  extending more  than
100~kpc to the SE of  M31's disk, stands  out from the background halo
not      only  in       stellar     over-densities     and   kinematics
\citep{ibata01a,ibata04,guhathakurta05b},  but  also   in  metallicity
\citep{ferguson02}.

Over  the next decade an immense  amount of data will become available
for stars   in the  Galaxy  from  both  astrometric  and spectroscopic
surveys     $-$  e.g.    from    the    astrometric  satellite    GAIA
\citep[e.g.][]{perryman01},  the ``Radial Velocity Experiment'' (RAVE)
\citep[e.g.][]{steinmetz03},  and  the ``Sloan  Extension for Galactic
Underpinnings                  and                      Exploration'',
SEGUE\footnote[1]{http://www.sdss.org}.  Positions,  line     of sight
velocities  and proper motions will  be measured for millions of stars
in      the   halo,    enabling  the   full       phase-space  to   be
reconstructed.  Similarly,  high accuracy, wide  field measurements of
stellar spectra will  allow  the mapping   of the  Galaxy  in chemical
abundances.  These studies will not only constrain cosmology and 
galaxy formation \citep[e.g.][]{bullock01} but also test ideas about
first light and reionization \citep[e.g.][]{tumlinson04}
With the upcoming observational data, the goal of putting
together the history of our Galaxy is finally becoming feasible. It is
therefore important to have theoretical  predictions for the  combined
kinematic  and chemical abundance distributions, as  well as to have a
coherent theoretical framework for interpreting the upcoming results.

Although  future surveys promise  to   provide  detailed maps of   the
chemical and phase-space  structure of the Galaxy,  we may worry  that
ours is but one among many galaxies  of its size.   How typical is our
galaxy?  How can it be used to constrain general ideas about cosmology
and galaxy formation  if it is  atypical in some way?  Fortunately, in
the context of  $\Lambda$CDM,  we have  well  defined expectations  for
variations in the formation times and accretion histories of Milky-Way
size galaxies and this idea is  at the heart  of our exploration.  For
example \cite{mouhcine05a,  mouhcine05b}  recently have  estimated the
metallicities of the inner halos of several nearby spiral galaxies and
find that they  are quite metal  rich compared to  the stellar halo of
the Milky Way.  They  go  on to suggest  that  the  Milky Way  is  not
typical for a normal  spiral galaxy of  its luminosity.   As discussed
below, if this result holds then our  models provide a straightforward
interpretation: that the Milky Way  has experienced fewer than average
major accretion events  with  LMC-size objects.   Indeed, phase  space
information could test  ideas of this  kind specifically by  revealing
few major accretion events and many more lower-mass accretions.

The  modeling of the  chemical evolution of  the Galaxy in a realistic
cosmological   context    is     just  beginning  to     be   explored
\citep[e.g.][]{bekki01,brook03,brook04}. Recent advances in   modeling
techniques, like the advent of ``hybrid'' methods, allow us to address
this   problem with  greater  resolution.  Hybrid  methods enable  the
modeling  of star formation   and  chemical enrichment  with  detailed
semi-analytical prescriptions,  and   at  the same  time  provide  the
detailed     dynamics     via     the coupled      N-body  simulations
\citep[e.g.][]{kauffmann99,springel01}.  In   this  study   we  use  a
previously developed hybrid  method \citep{bullock05,robertson05,font05}
to explore the  phase-space distribution of  chemical abundances  in a
series  of  Milky  Way-type galaxy halos    formed in  a  $\Lambda$CDM
Universe   ($\Omega_{m}=0.3,  \, \Omega_{\Lambda}=0.7, \,  h=0.7,$ and
$\sigma_{8}=0.9$).

A description of our sample of stellar galaxy models is given in Section \ref{sec:sample}. In Section \ref{sec:results} we present results on the spatial distribution of \FeH and \alphaFe chemical abundances in these stellar halos and propose a criterion for detecting cold stellar streams based on their \alphaFe abundances. In Section \ref{sec:comparisons} we discuss our results and compare them with the available observations and in Section \ref{sec:conclusions} we conclude.

\section{The Sample of Milky Way-type Stellar Halos}
\label{sec:sample}

In a previous  study \cite{bullock05}  have presented  a  sample of 11
stellar halos of Milky Way-type galaxies assembled hierarchically from
accreted satellite galaxies in a
$\Lambda$CDM cosmology. The physical properties of the simulated halos
are summarized  in Table~1.  The  merger histories of these halos have
been specifically selected not to have high mass recent mergers, so as
to maximize the probability that it will host a disk galaxy like the
Milky Way \citep[e.g.][]{wyse01}.  
Our model stellar halos have  similar
mass, density  profile and total  luminosity as the Galactic halo. The
models  also  obtain roughly  the same  number  of surviving satellite
galaxies as in the Milky Way and match  their physical properties: the
stellar mass-circular velocity relation, $M_{*} - $v$_{\rm circ}$, for
Local Group dwarfs; and  the distribution  of surviving satellites  in
luminosity,   central     surface  brightness  and   central  velocity
dispersion.

In this paper we  extend our models to include
chemical evolution with the prescriptions of \cite{robertson05}.
These include enrichment from both Type II  and Type Ia supernovae
and  feedback  provided by supernovae    blow-out and winds  from
intermediate mass stars.  We  follow the  chemical evolution of   each
satellite galaxy accreted onto the  main Galaxy and trace the build-up
of a  series of  metals, such  as  Fe or $\alpha$-elements.  Each star
particle in the simulations  has assigned a set  of \FeH and  \alphaFe
abundance{\footnote [2]{The  \alphaFe ratio is  defined as the average
of  [Mg/Fe]  and [O/Fe] ratios,  and  used thereafter  in the paper.}}
ratios  determined by the star formation   history of the satellite in
which  the star  originated.   The coupled N-body  simulations resolve
each Milky  Way-type stellar  halo with  a few million  particles  and
allow  us    to explore  the   phase-space   distribution of  chemical
abundances in unprecedented detail.

\begin{table}[]
\resizebox{!}{3cm}{\begin{tabular}{ccccccc}
\hline
			    	&  \#	   		& \#   		& $M_{*}^{halo}$  	& 80\% halo 	&	time of  	& Number of \\
Halo	   	& accreted	& surviving   	& 	(<300kpc)     	& accretion 	& last $>$10\%  & particles \\
	 	& luminous	&  luminous 	& 				& time 		&  merger &  in stellar\\
	   	& satellites	 		&  satellites	&	($10^9 M_\odot$)	& (Gyr)		& (Gyr)	& halos \\
\hline 
H1   		&  115 		& 18    		& 	4.16		 	& 	5.3		&  8.3 	& 	1908100 \\
H2   		&  102 		& 6	    		& 	3.26		 	&	7.0		&  9.2	& 	1890197\\
H3   		&  106 		& 16	    		& 	4.17		 	&	7.4		&  8.9  	&  	1689035\\
H4   		&  97	 		& 8	    		& 	4.11		 	&	6.3		&  8.3 	& 	1628686\\
H5   		& 160 		& 18	    		& 	2.40		 	&	2.1		& 10.8 	&	2531076\\
H6   		& 169 		& 16	    		& 	2.45		 	&	6.2		& 10.5  	& 	2873983\\
H7   		& 102 		& 20	    		& 	2.44		 	&	4.4		&  7.4 	&	1797682\\
H8   		& 213 		& 13	    		& 	2.58		 	&	7.1		&  9.3 	& 	4246733\\
H9   		& 182 		& 15	    		& 	2.52		 	&	1.5		& 10.0 	&	2528762\\
H10  		& 156 		& 13	    		& 	3.12		 	&	2.9		& 9.7  	& 	2785418\\
H11  		& 153 		& 10	    		& 	2.56		 	&	7.2		& 9.0	         &	2598580\\
\hline
\end{tabular}}
\caption{ \label{halo_tab} Properties of the simulated stellar halos.Times refer to lookback times.}
\end{table}

\section{Results}
\label{sec:results}

\subsection{Global Properties of the Simulated Halos}
\label{sec:global}

Tidal streams  in the  halo    display  various  degrees  of   spatial
coherence,  given  their  different  times of  accretion  and  orbital
properties. Similarly,  we expect to  see various degrees of coherence
in  the chemical abundance space,  given the  different star formation
and chemical enrichment histories of their progenitor satellites.

\begin{figure}
\figurenum{1}
\epsscale{1.2}
\plotone{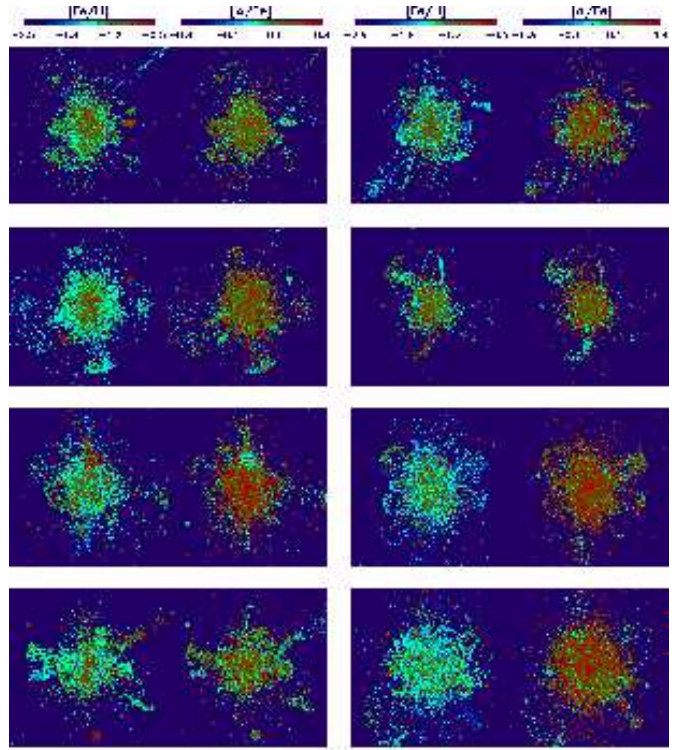}
\caption{\label{fig:metal_map_halos_feh_afe}{$(x,z)$ projections of chemical abundances in halos H1 $-$ H8 (in order from top left to bottom right).  For each halo we show side by side the star particles color-coded in \FeH (left sub-panels) and in \alphaFe (right sub-panels). The maps span 300~kpc on the size. Each pixel in these maps extends 0.25~kpc in the $x$ and $z$ directions and 300~kpc in the $y$ direction (i.e., in depth). The \FeH and \alphaFe values are mass weighted averages over all stars contained in each pixel. 
}}
\end{figure}

Figure \ref{fig:metal_map_halos_feh_afe} shows the present day ($t=0$)
spatial  distribution of \FeH and  \alphaFe abundance ratios for halos
H1 $-$ H8. The maps span 300~kpc on a side  ($x$ and $z$ directions,
respectively) and 300~kpc  in  depth  ($y$). Each pixel  represents  a
color coded chemical   abundance, calculated as  the  average  \FeH or
\alphaFe  of all  stars included in  a  volume $0.25 \times 300 \times
0.25$~kpc$^{3}.$  Our models  show  that  inner  regions of the  halos
($r<50$~kpc) are consistently   more homogeneous than the   outer ones,
both in terms of the spatial distribution  of stars and in the overall
chemical abundances. This is because the inner halo builds up rapidly,
with most of its stellar mass already in place  in the first $4-5$~Gyr
of the Galaxy evolution \citep{bullock05,font05},  and at these  times
the  dynamical   timescales  were   shorter   and   phase-mixing  more
efficient. The spatial distribution of the stars has been discussed in
more detail by  \cite{bullock05}, who also  note a similar behavior in
the surface brightness  and velocity distribution  of halo stars. Here
we concentrate on the corresponding \FeH and \alphaFe distributions.

We find that  the inner  $\sim 50$~kpc regions  are,  on average, metal
enriched (\FeH $\ge -1.2$) and we trace this to the metal contribution
of  a  few  massive ($M_{*}   \sim  10^{9} \,   M_{\odot}$) satellites
accreted early on and which  contribute a significant fraction ($40-60
\%$) to the total stellar mass  of the halo \citep{font05}.  The rapid
growth  of these  satellites at early   times  fueled an  intense star
formation and therefore  at the time  of accretion the satellites were
already   enriched in \FeHt.  In  the  outer regions ($r>50$~kpc)  the
streams are more  spatially   separated  and their  various   chemical
enrichments are more distinct.  Both metal poor and metal rich streams
are present, with  the more metal  poor ones originating in satellites
with  low   star  formation   rates (these are    typically   low mass
satellites).  A  more  detailed discussion of  radial distributions of
stars will be given in \S \ref{sec:radial}.

\begin{figure}
\figurenum{2}
\epsscale{1}
\plotone{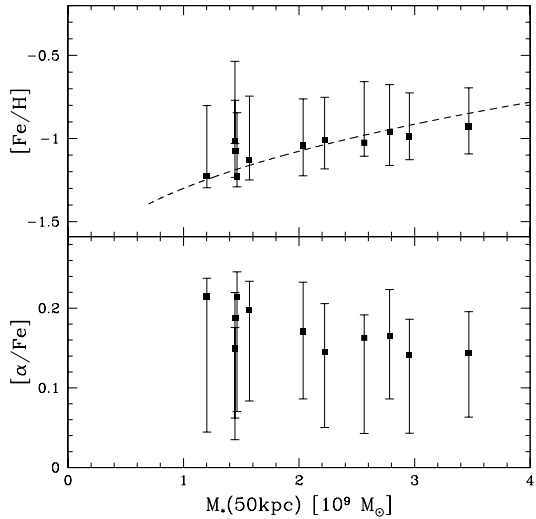}
\caption{\label{fig:metal_halo_averages}{Mass weighted  \FeH and \alphaFe average values versus stellar mass for halos H1 $-$ H11.   Averages are calculated within $r<50$~kpc, a cutoff value which approximates the inner few tens of kpc currently probed by observations. The dashed line in the top panel corresponds to the \FeH $\sim M_{*}^{2/5}$ fit of \cite{dekel03}. Errorbars represent weight averaged $25 \%$ and $75 \%$ values of the absolute spread in abundance ratios. }}
\end{figure}

Figure   \ref{fig:metal_halo_averages}  shows  the   average \FeH  and
\alphaFe for all  our Milky Way-type halos.  The general trend in  our
simulations  is  that the    more  massive a   stellar halo,  the more
chemically evolved it is, i.e. the higher \FeH  and lower \alphaFe it
has (these
relations  are   independent   of  the  cutoff  radius    assumed  for
$M_{*}(r)$).       The  \FeH   $-         M_{*}$  trend   in    Figure
\ref{fig:metal_halo_averages} is consistent with having the
same slope as that of   the   well-known  relation between
metallicity and luminosity seen in low-mass galaxies, 
\FeH $\propto M_*^{-2/5}$ (dashed  line).
We note that our  chemical feedback model is normalized such that
surviving satellites at $z=0$ match 
 the observed metallicity-luminosity relation for Local Group galaxies
\citep{larson74,dekel86,dekel03}.  It  is interesting  however that
a similar relation propagates to the stellar halos themselves.
This follows from the fact that the slope of the relation for
satellite galaxies is governed
by their parent halo potential well depths.  Therefore 
 disrupted satellites also
follow a \FeH  $\propto M_*^{-2/5}$ relation 
\citep[see Figure 4][]{font05}, but with
a lower \FeH normalization because these  systems tend to
be accreted earlier and are  less evolved \citep{font05,robertson05}.
The same relation propagates to a {\em stellar halo} mass-metallicity
relation because stellar halo mass correlates strongly with the number
of disrupted {\em massive} satellites (see \S \ref{sec:mah}).  
The more massive disruptions a stellar halo
contains, the more metal rich the stellar halo becomes.  Similarly,
because its progenitors
are massive, they are more metal rich, and the halo is higher metallicity
as a result.

Our  results suggest that  the  hierarchical assembly is an  important
factor in determining the metallicity - mass relation for stellar halos.
A similar argument can be extended to qualitatively understand the 
observed trend presented by \cite{mouhcine05b}, who show that low luminosity
spiral galaxies have more metal poor central stellar halos than do high luminosity
spirals.  This qualitative trend can be explained in a hierarchical scenario:
low-luminosity spirals tend to have accretion histories dominated by correspondingly
lower-luminosity dwarfs.  Lower luminosity dwarfs are more metal poor and thus
produce lower metallicity stellar halos.
  While one   can
derive the  \FeH $- M_{*}$ relation based  on simple scaling arguments
that  relate  the metal  loss  to the depth of  the  central galaxy's
dark mater halo  \citep[e.g.][]{larson74,tinsley80}, this does
not necessarily imply that the relation forms in situ.

We caution that
for a complete  understanding of the origin of  the relation  (both of
its shape  and normalization), one  needs a more accurate treatment of
gas physics  processes in satellites after they  are accreted onto the
main halo. These effects are not currently included in our models, but
we  expect  them to   be of  second  order   to the  hierarchical mass
assembly.

\subsection{Mass accretion histories and halo metallicities} 
\label{sec:mah}

We can verify the hierarchical origin of the \FeH $- M_{*}$ relation by investigating the mass accretion histories of the stellar halos. Given that in a typical mass accretion history the few massive satellites contribute significantly to the final stellar mass, we expect the overall halo metallicity (and hence the \FeH $- M_{*}$ trend) to be determined mainly by these massive accretions. The time of accretion of massive satellites is also a related factor $-$ satellites accreted later are likely to be more metal rich. 

\begin{figure}
\figurenum{3}
\epsscale{1}
\plotone{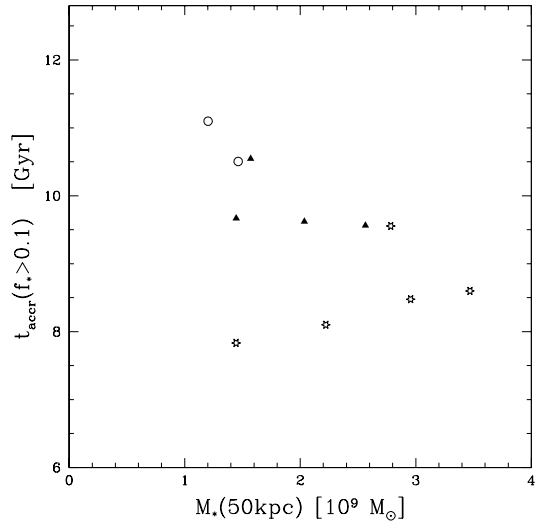}
\caption{\label{fig:metal_halo_averages_taccr}{The lookback time of accretion of the most massive satellites, calculated as the average $t_{accr}$ of those satellites which contribute $\ge 10 \%$ to the total mass of the stellar halo) versus stellar mass of the inner halo, $M_{*}$(50kpc). Star symbols correspond to the most metal rich halos, triangles to halos of intermediate \FeH and circles to those of lowest \FeH in our sample.}}
\end{figure}

Figure \ref{fig:metal_halo_averages_taccr} plots for each stellar halo
in our sample   the lookback time  of accretion  of  the  most massive
progenitors, specifically    an average  over those   satellites which
contribute more than 10$\%$ to the total stellar fraction. The general
tendency   is    that     more    massive   halos    assemble     more
recently. Additionally, for halos of the same  mass the spread in \FeH
can be also explained by differences in their formation time (e.g. the
three   halo    models   with    $M_{*}   \sim  1.5     \times  10^{9}
M_{\odot}$). Therefore a  protracted  assembly   of the higher    mass
stellar halos seems to be at the origin of the \FeH $- M_{*}$ relation
\citep[see also similar results of][]{renda05}.

\vskip 6pt
\noindent {\it A Test Case:  Milky Way and M31}

We further  illustrate the  relation between mass  accretion histories
and    the  overall metallicities   with the  case  of  Milky  Way and
M31. These are two large  galaxies with  roughly the same  luminosity,
but the Milky Way halo is more metal poor than that of M31, perhaps by as
much as  1~dex.   
The metallicity distribution  function (MDF)  in the Milky Way
halo peaks around  \FeH $= -1.5$ \citep{laird88,carney96}, whereas the
MDFs  measured  at several locations  in  the {\it inner} ``halo'' of
M31 peak  at values
between \FeH $\simeq -0.4$ and $-0.7$ \citep{reitzel02,bellazzini03}.
Although, as discussed below, 
the outer halo of M31 at \FeH $\sim -1.2$ may be a more realistic
estimate of the true underlying 
stellar halo component of this galaxy \citep{guhathakurta05a}.

\begin{figure}
\figurenum{4}
\epsscale{1}
\plotone{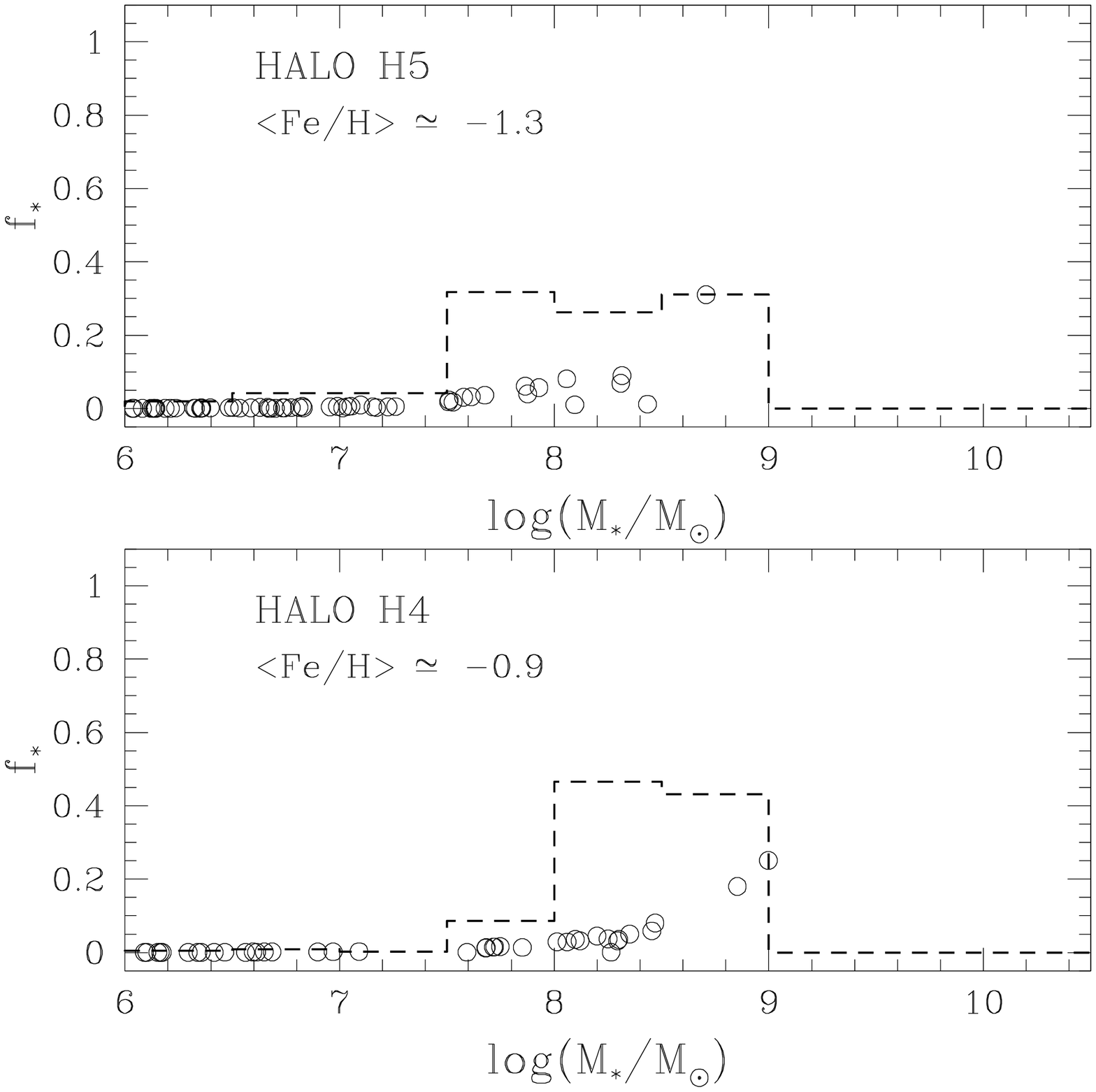}
\caption{\label{fig:mah_halos_08_09}{Mass accretion histories expressed as the stellar fraction $f_{*}$ of the inner ($r<50$~kpc) halo versus the stellar mass $M_{*}$ of the contributing satellites. The top panel corresponds to halo H5, a metal poor halo with \avFeH $\simeq -1.3$, and the bottom panel to halo H4, a metal rich halo with \avFeH $\simeq -0.9$. Empty circles show individual contributions of satellites to the stellar fraction, and dashed lines their cumulative contribution. 
}}
\end{figure}

Figure  \ref{fig:mah_halos_08_09} shows  the mass  accretion patterns,
expressed as the fractional contribution of accreted satellites to the
total stellar mass of the  inner halo, for  two halo models similar to
inner Milky Way and M31: halo H5, with  \avFeH $ \simeq  -1.3,$ and halo H4,
with \avFeH $ \simeq  -0.9$ (these are also  the most extreme cases in
our  sample).  The more  metal poor  halo has  one  major accretion of
stellar mass, $M_{*} \sim 10^{8 -  9} M_{\odot}$ accreted about 11~Gyr
ago, while the more  metal rich one has two  major accretions, both of
masses $M_{*} \simeq   10^{9}  M_{\odot}$ and  accreted  about 8.5~Gyr
ago. Thus  the most massive, metal rich  halo had  an assembly history
which includes more massive progenitors and which were accreted later.

The  inference from  these    results   is  that the     difference in
metallicities  between   the two galaxies   can   be explained if  M31
experienced one or  two more  massive  accretions than the Milky   Way
and/or a protracted merger history.

However  we note that the metallicity  measurements in  both Milky Way
and M31 are  not  yet completely reliable.  One  reason is that   most
observations probe  only the inner few  tens of kpc  and therefore may
not be representative of the  entire halos  or may include  additional
galactic components. For example, recent observations in M31 extending
further out into  the halo find  \FeH $\simeq  -1.2$ for stars  beyond
60~kpc, suggesting that  the high metallicities  measured in the inner
regions     include the contribution      from  M31's  extended  bulge
\citep[][Gilbert    et~al.,    in      prep;]{guhathakurta05a,kalirai05}.  
Additional  contamination may  come from the
recently discovered disk-like component extending out to $\sim 70$~kpc
\citep{ibata05,irwin05}. There are also differences in the measurement
methods.    In    the   Milky  Way    metallicities   are   determined
spectroscopically,  while   in  M31   they   are  mostly    determined
photometrically \citep[but    see  new  spectroscopic     measurements
of][]{ibata04,guhathakurta05b}.   It is therefore  unclear whether the
metallicity  discrepancies  between  the  two  galaxies are  real, but
ongoing wide-field spectral studies in these galaxies may be soon able
to answer this problem. Nevertheless, our results are more general and
suggest ways  in   which massive  accretion   events can increase  the
average   metallicities of halos  (eg. 1  massive dwarf galaxy, $M_{*}
\simeq  10^{9}  M_{\odot}$,  accretion is  sufficient to  increase the
average metallicity of a Milky Way-type halo by about 0.5~dex).

\subsection{Radial Distributions}
\label{sec:radial}

As explained in the Introduction, metallicity gradients are a powerful
tool  for  constraining the timescale  of formation   of galaxies.  An
often-cited hypothesis is that hierarchical formation will result in a
negligible metallicity gradient.   In detail, however, some variations
may occur, particularly in  regions  which continue to  assemble until
recent times.

\begin{figure*}[t]
\figurenum{5}
\epsscale{1}
\plotone{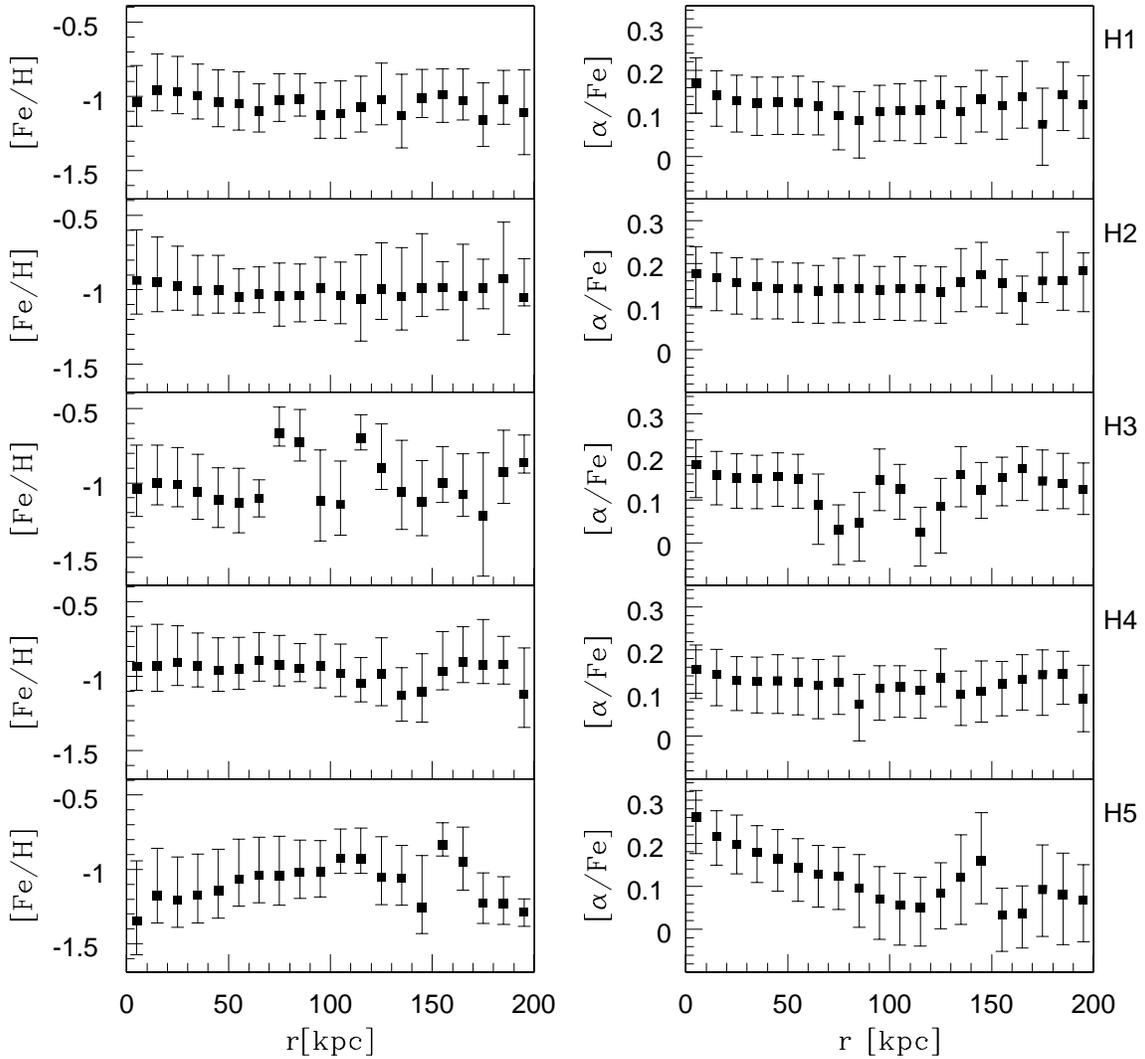}
\caption{\label{fig:metal_gradient_1}{Radial distributions of \FeH (left panels) and \alphaFe (right panels)  for halos H1 $-$ H5. The chemical abundances are weight averages in radial shells of width $dr=10$~kpc. Error bars represent weight averaged $25 \%$ and $75 \%$ values of the absolute spread in abundance ratios within each radial bin.
}}
\end{figure*}

\begin{figure*}[t]
\figurenum{6}
\epsscale{1}
\plotone{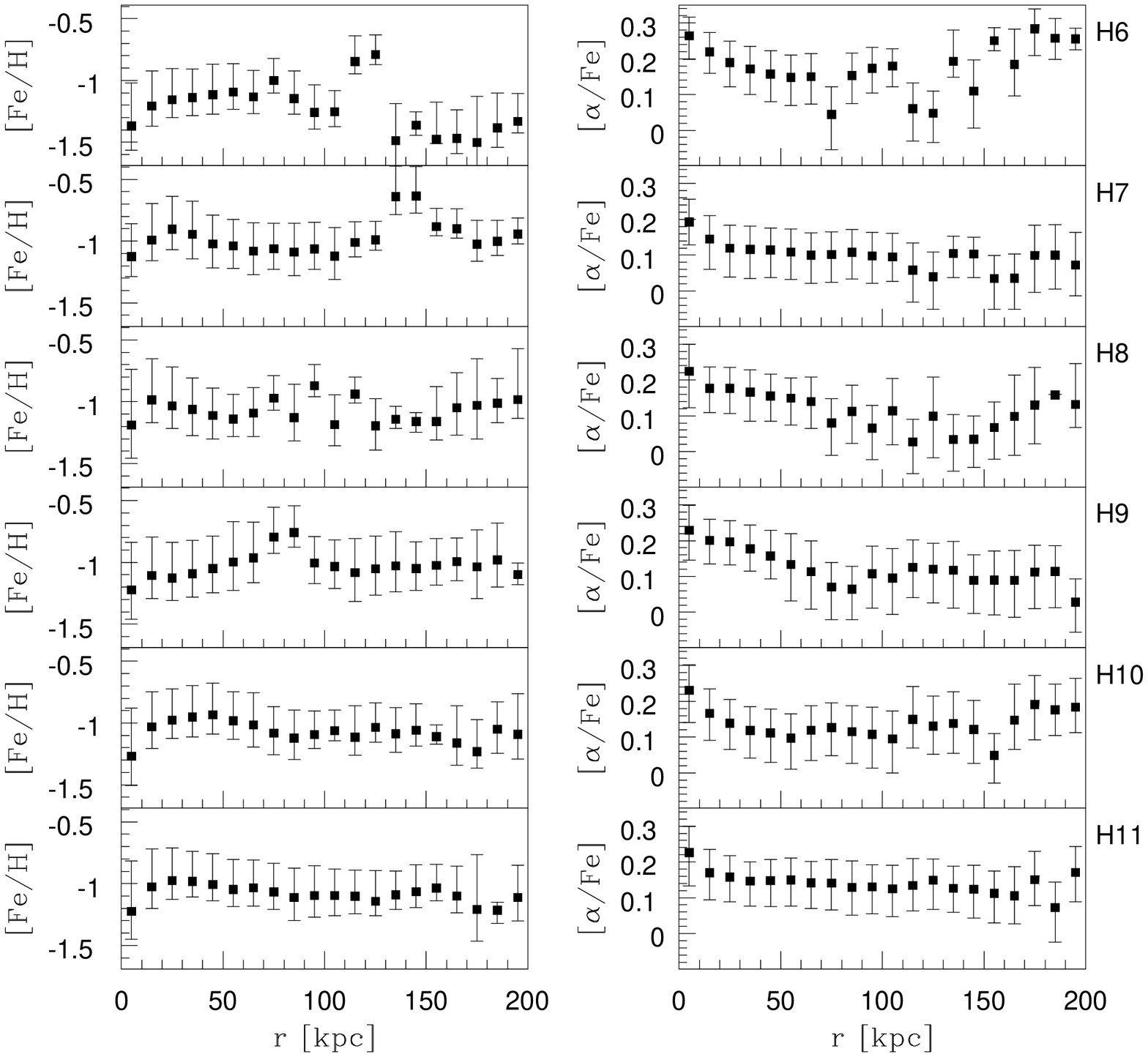}
\caption{\label{fig:metal_gradient_2}{Radial distributions of \FeH (left panels) and \alphaFe (right panels)  for halos H6 $-$ H11. The abundance ratios are weight averaged values within radial bins of 10~kpc. Error bars represent weight averaged $25 \%$ and $75 \%$ values of the absolute spread in accretion times and ages within each radial bin.
}}
\end{figure*}

The radial distributions of  \FeH   and \alphaFe abundances for    all
simulated halos    are shown Figures    \ref{fig:metal_gradient_1} and
\ref{fig:metal_gradient_2}. The  distributions do not show significant
global gradients, but many do show variations over scales
of a few tens of kpc.

\begin{figure*}[t]
\figurenum{7}
\centerline{\epsfysize=6.truein \epsffile{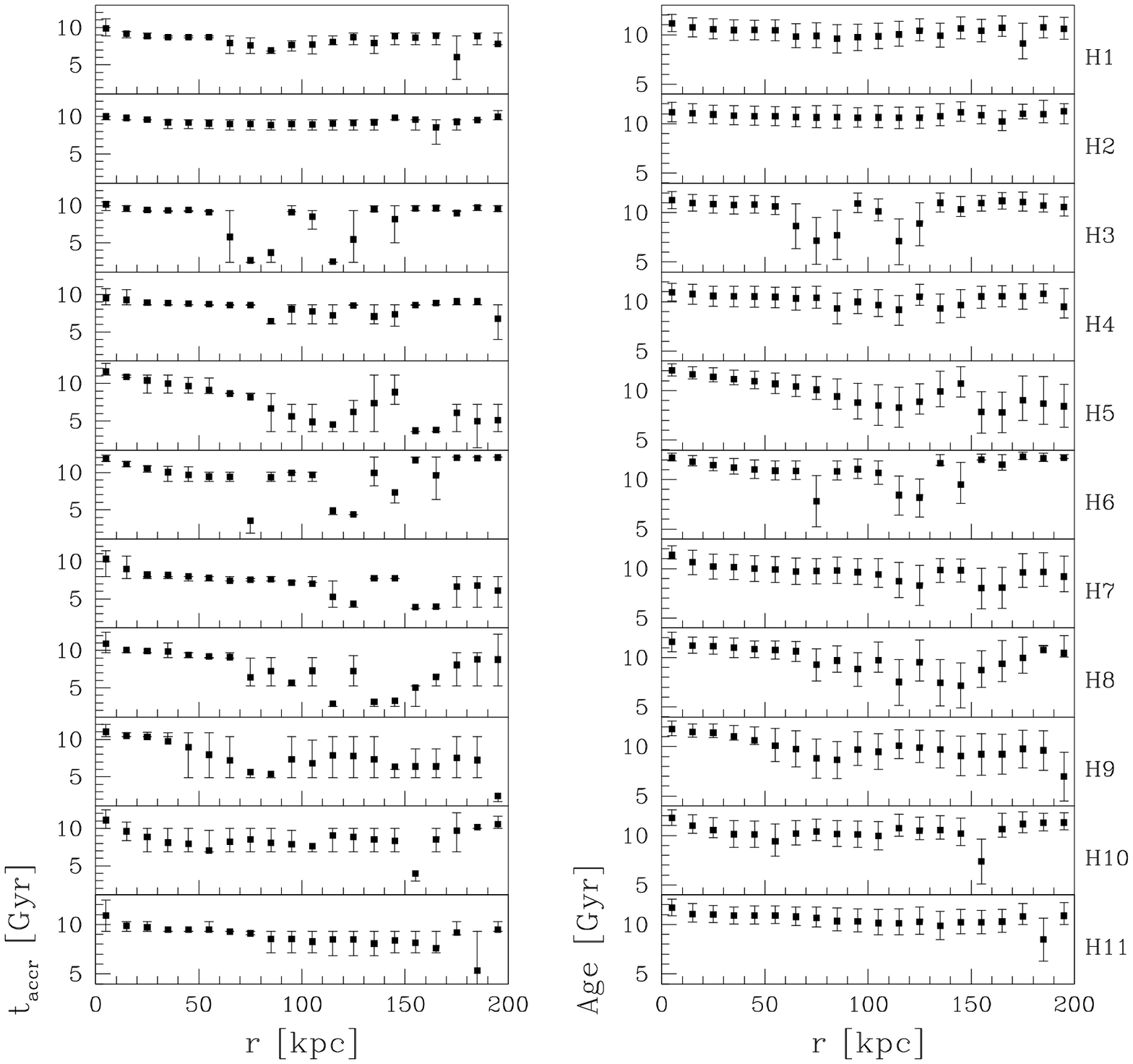}}
\caption{\label{fig:metal_time_gradient}{Radial distributions of lookback accretion times (left panels) and ages of stellar populations (right panels)  for halos H1 $-$ H11. The abundance ratios are weighted averages within radial bins of 10~kpc. Errorbars represent weight averaged $25 \%$ and $75 \%$ values of the absolute spread in abundance ratios in each radial bin.
}}
\end{figure*}

The lack of large-scale gradients can be explained by 
examining the mass  accretion  patterns  of our  simulated  halos.  The bulk  of the
stellar halos forms  early, both in the inner  and outer regions.  The
outer  halos continue to accumulate  stellar debris  up to more recent
times ($\sim 5-6$~Gyr ago), but this  represents only a small fraction
of the total stellar mass \citep{bullock05,font05}. Therefore, even if
the later assembly  of the outer halo favors  an influx  of chemically
enriched   material,  this is  not   sufficient to  create significant
large-scale gradients. Moreover,  the chemical abundances  are directly correlated
with   the ages   of   the  stars, rather    than   with the time   of
accretion. Figure \ref{fig:metal_time_gradient} plots the distribution
of   lookback accretion times (left panels)   and the  ages of stellar
populations (right  panels) versus the radial   distance, for all halo
models. As this figure shows, even if the stellar material is accreted
recently,  it usually contains a  mixture of both  young and old stars
that dilutes the metallicity differences.

Many of our halos do show distinct small-scale
 abundance  gradients. For example, in  halos
H5, H6 and H9 the average \FeH increases with $r$ in the inner 100~kpc
(similarly, \alphaFe decreases  over the same distance). The gradients
in these halos  arise from the  spatial distribution of their  stellar
populations. Figure \ref{fig:metal_time_gradient} shows that the inner
regions of these    three  halos contain mostly  old   ($\sim 10$~Gyr)
stellar populations, whereas the outer regions have on average stellar
populations a few Gyr younger.  This result has implications for 
searches for the lowest metallicity stars in the stellar halo and
complimentary attempts to use the metallicity distribution function
to constrain the nature of reionization and PopIII star formation
\citep[e.g.][]{tumlinson04,beers05}.  Specifically, the lowest metallicity
stars are expected to inhabit the most central regions of the
Galaxy (note the inner-most radial bin in Figure \ref{fig:metal_gradient_2}).  
Thus general attempts to use observed metallicity distribution
functions to constrain early chemical enrichment must take special
care to account for spatial biases in observational samples of stars.

The  radial  distributions of  accretion   times,  ages and   chemical
abundances give an insight into the composition of stellar populations
in  the simulated halos.  Our results  suggest that  stellar halos  of
galaxies like the  Milky Way should  contain a  large fraction  of old
($>10$~Gyr) stellar  populations which were accreted  in the first few
Gyr of the mass assembly. In addition  to the underlying population of
old  stars, some  halos contain small   fractions  of intermediate age
($\sim 7-10$~Gyr) populations. Irrespective of their age, stars can be
either  metal  poor  or metal rich,  depending  of  the mass  and star
formation rate of their progenitor satellite.

Interestingly, in terms of its age  and metallicity distributions, the
Milky  Way halo lies  at the edge  of our  sample simulated halos. The
stars in the Milky Way halo are mostly  old and metal poor, whereas in
our models   a significant  number  of old  stars  are  already  metal
enriched. (Note  however that the typical  observations cover only the
inner $\sim 10-20$~kpc of the halo and in this range some of our halos
are old and metal  poor). Our results seem to  suggest that  the Milky
Way  halo is  an  atypical case in   the range of  galaxies of similar
mass. The metallicity  constraints rule out  an accretion of a massive
($M_{*} \sim 10^{9} M_{\odot}$)   halo progenitor at early times  when
the  bulk  of the halo formed.  We  also note that recent observations
show that many other bright spiral  galaxies of similar mass have more
metal  rich   halos  than that of  the   Milky  Way. A   more detailed
discussion   of  these observations  will     be given in Section   \S
\ref{sec:feh}.

\subsection{Mining the Halo for Stellar Streams}
\label{sec:sig}

Tidal streams are fossil records of  past accretions and therefore can
be  used to  reconstruct  the   merger  histories of   galaxies. Their
detection however is difficult due to their low surface brightness and
to the loss in spatial coherence the longer they  orbit in the Galaxy.
The identification of stellar streams can be improved by extending the
parametric   space to include    new   dimensions like the   kinematic
parameters and chemical abundances.

\subsubsection{Kinematics}
\label{sec:kin}

If kinematic data is available from observations, one can use the $(E,
\, L_{z}, \,  L)$ space to  identify stellar  streams. This parametric
space is  particularly useful for epochs  of slow evolution,  when the
time required for stars to exchange energy and angular momentum is long
compared with the age  of the Galaxy, and  $E$, $L_{z}$ and $L$ can be
approximated as invariants of motion \citep[e.g.][]{eggen62}.  In this
case, the tidal debris from different  accretions is expected to clump
in  distinct locations in      this space. Numerical  simulations   of
satellite accretions in a fixed potential  confirm that this is indeed
the case \citep{helmi00}.

\begin{figure}
\figurenum{8}
\epsscale{1.2}
\plotone{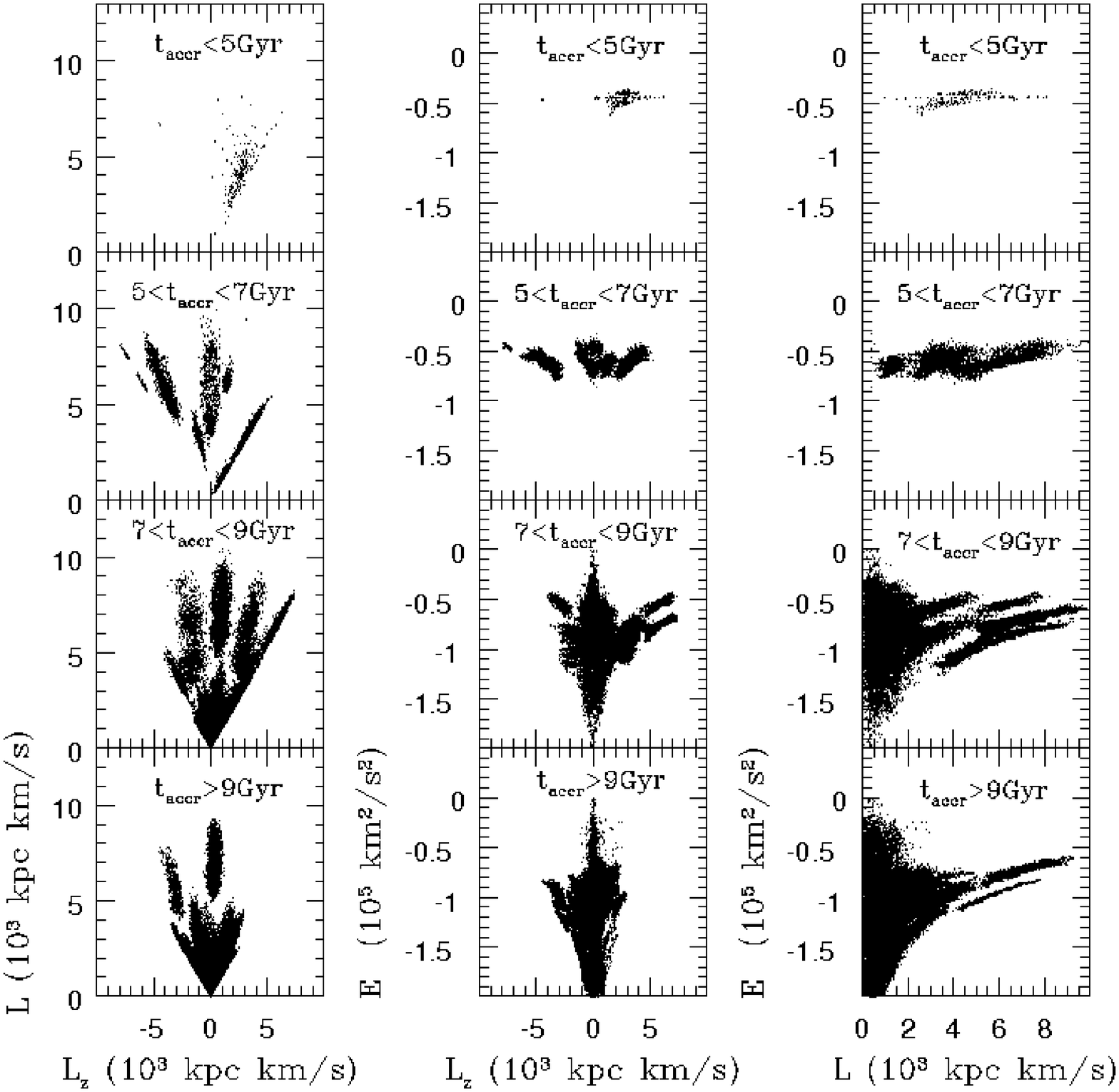}
\caption{\label{fig:metal_en_angmom_halo02}{Distribution of stars in halo H1 in the $L_{z} - L$ space (left panels),  in $E - L_{z}$ (intermediate panels) and in $E - L$ (right panels).  Different panels from top to bottom correspond to diffferent lookback accretion times of the satellites where stars originate from.
}}
\end{figure}

In contrast with previous studies we  model the mass accretion history
in a cosmological context  and  include a time-evolving  gravitational
potential which  grows  with the  mass of the  Galaxy  \citep[see][for
details]{bullock05}. We   want to see if,    with these additions, the
tidal   debris can still  be  recovered in  the $(E, \,  L_{z}, \, L)$
space.  Figure \ref{fig:metal_en_angmom_halo02} plots the distribution
of halo H1 stars in the $L_{z} - L$, $E - L_{z}$ and $E - L$ planes at
different   ranges of  lookback   accretion  times of the   progenitor
satellites.  This   shows that the   recent accretions  ($t_{accr} \le
7-9$~Gyr) occupy relatively   distinct locations in these  planes  and
their identification is therefore possible.

The clumps  in   the low $E$  and  $L$  regime are more   difficult to
separate,  in part  because of their  superposition,  but also because
this is also the regime of rapid growth  of the galaxy when the energy
and angular momentum of stars change significantly. Additional effects
like  the interaction  between   satellites  and the  non-axisymmetric
oscillations in the  Milky  Way potential,  not modeled in  our study,
will further disperse the stars in phase-space \citep{mayer02}.

The kinematic identification of  stellar streams in the halo  requires
high accuracy  measurements of proper motions  over extensive areas of
the sky.   Future  astrometric Galactic surveys such  as  GAIA will be
able to  provide   these  data and   identify  some  of  these  clumps
\citep{helmi00}.

 \subsubsection{Chemical Abundances}
 \label{sec:afecut}

Chemical abundances can be used as an additional dimension to the $(E,
\, L_{z}, \,  L)$ parametric space to investigate  the origin of stars
\citep[e.g.][]{bekki01,dinescu02}.   By separating   the     available
kinematic   parameters    in    different   metallicity   ranges, some
observational studies  have found stars which stand  out from the rest
of         the       halo,         implying      a     merger   origin
\citep[e.g.][]{carney96,majewski96}. However, given that even    stars
from the smooth  halo may originate  in mergers, it  is instructive to
investigate   whether    chemical abundances   can   provide  any  new
information about the merger history of the galaxy.

\begin{figure}
\figurenum{9}
\epsscale{1}
\plotone{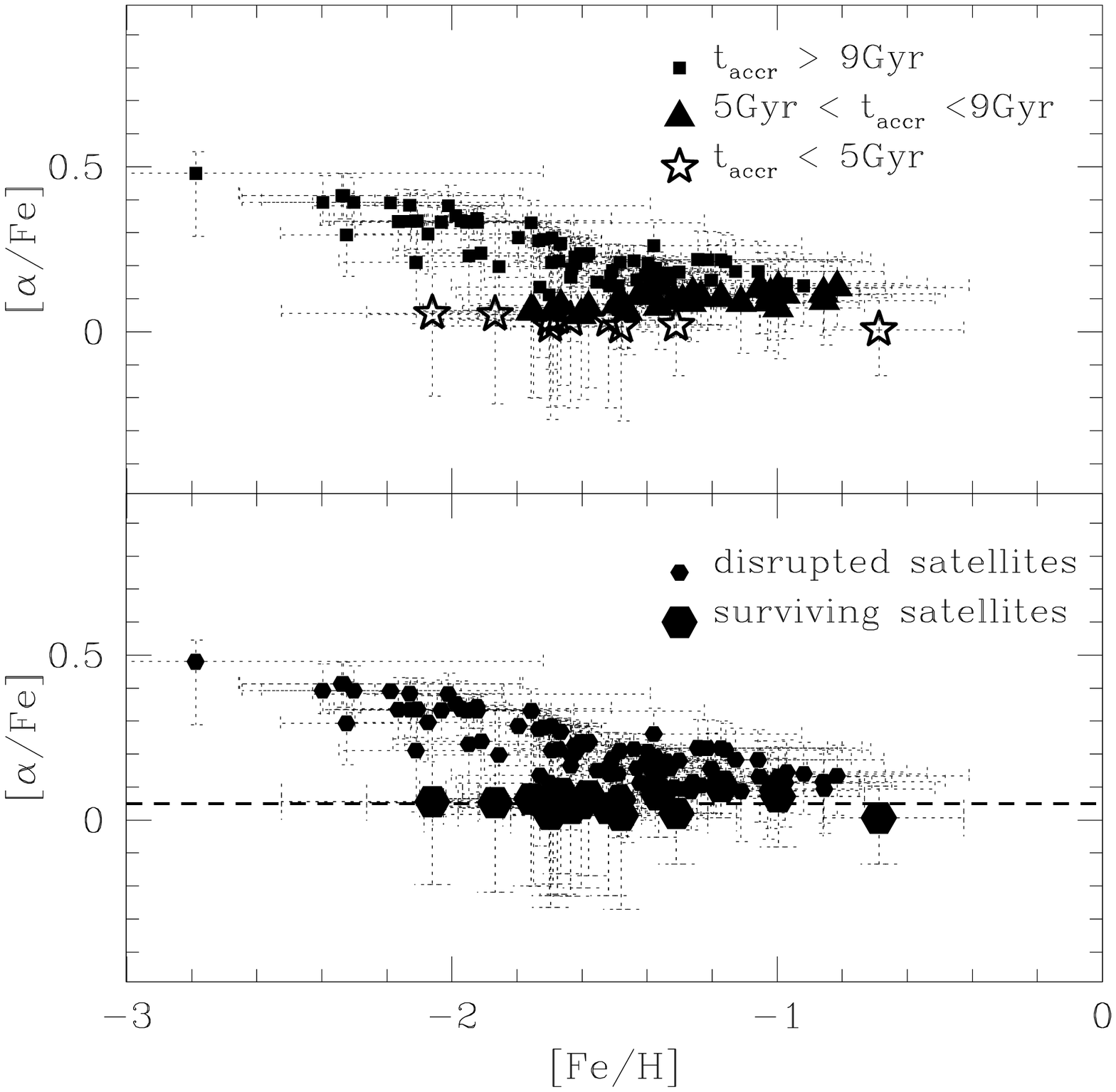}
\caption{\label{fig:average_afe_feh}{\alphaFe versus \FeH for all baryonic satellites accreted onto halo H1. {\it Top:} Satellites are plotted with different symbols function of their lookback time of accretion: squares denote satellites accreted at $t_{accr} > 9$~Gyr, triangles those accreted $5 < t_{accr}<9$~Gyr; and star symbols those accreted $t_{accr}<5$~Gyr. {\it Bottom:} Small and large symbols denote whether satellites are fully disrupted or still containing some bound material at present time $(t=0)$, respectively.  Errorbars in both represent $10\%$ and $90\%$ of the weight averaged values. 
}}
\end{figure}

\begin{figure}
\figurenum{10}
\epsscale{1.2}
\plotone{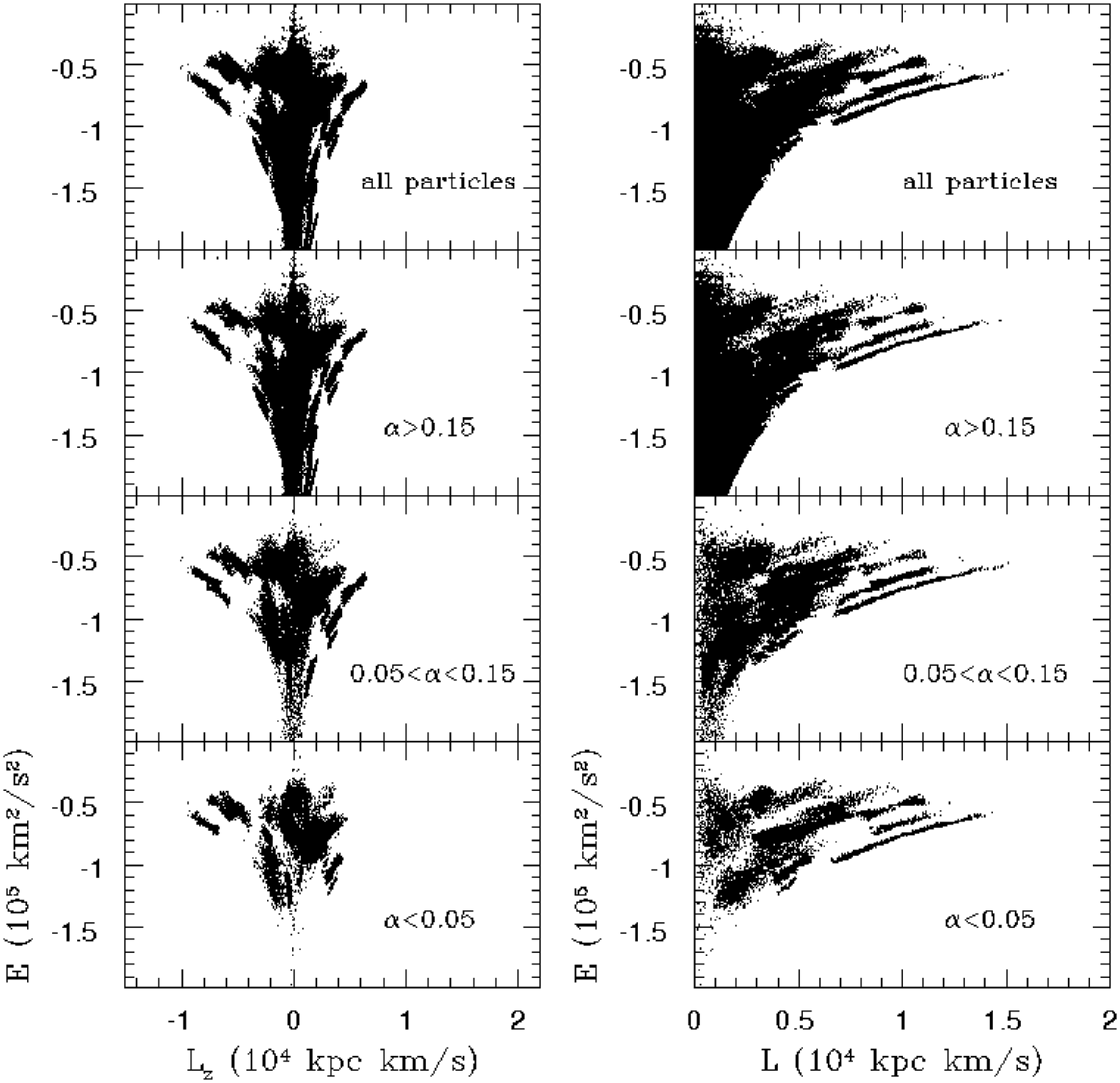}
\caption{\label{fig:metal_en_angmom_chem_halo05}{Distribution of stars in halo H2 in the $E - L_{z}$ space (left panels) and in $E - L$ (right panels). Top sub-panels show all stars, while the remaining three sub-panels from top to bottom correspond to different \alphaFe ranges.
}}
\end{figure}

In particular, chemical abundances may be able to constrain the merger
timeline of the Galaxy. In Figure \ref{fig:metal_en_angmom_halo02}, we
have used the time of accretion to separate the  clumps in the $(E, \,
L_{z},  \, L)$  space.  This parameter is  not easily  measurable, but
fortunately, we can use the natural clock provided by the evolution of
Fe and $\alpha$-elements. Each satellite  galaxy is expected to have a
distinct \alphaFe  -   \FeH evolution, however  it  has   some generic
features  such as a plateau  at high \alphaFe and  low \FeH and then a
steady decrease towards high \FeH (both the  length of the plateau and
the slope of the decrease being modulated by  the star formation rate,
hence the  mass of the  satellite). The plateau  is  maintained by the
enrichment in   $\alpha$-elements in Type  II supernovae  operating on
short timescales (of a few Myr) and the \alphaFe decrease is driven by
the later enrichment in Fe produced  in Type Ia supernovae.  Prolonged
episodes of  star formation will tend to  decrease the \alphaFe ratios
of stars in a satellite. This suggests an association between the time
of accretion of satellites and their overall \alphaFet.

\begin{figure}
\figurenum{1}
\epsscale{1.2}
\plotone{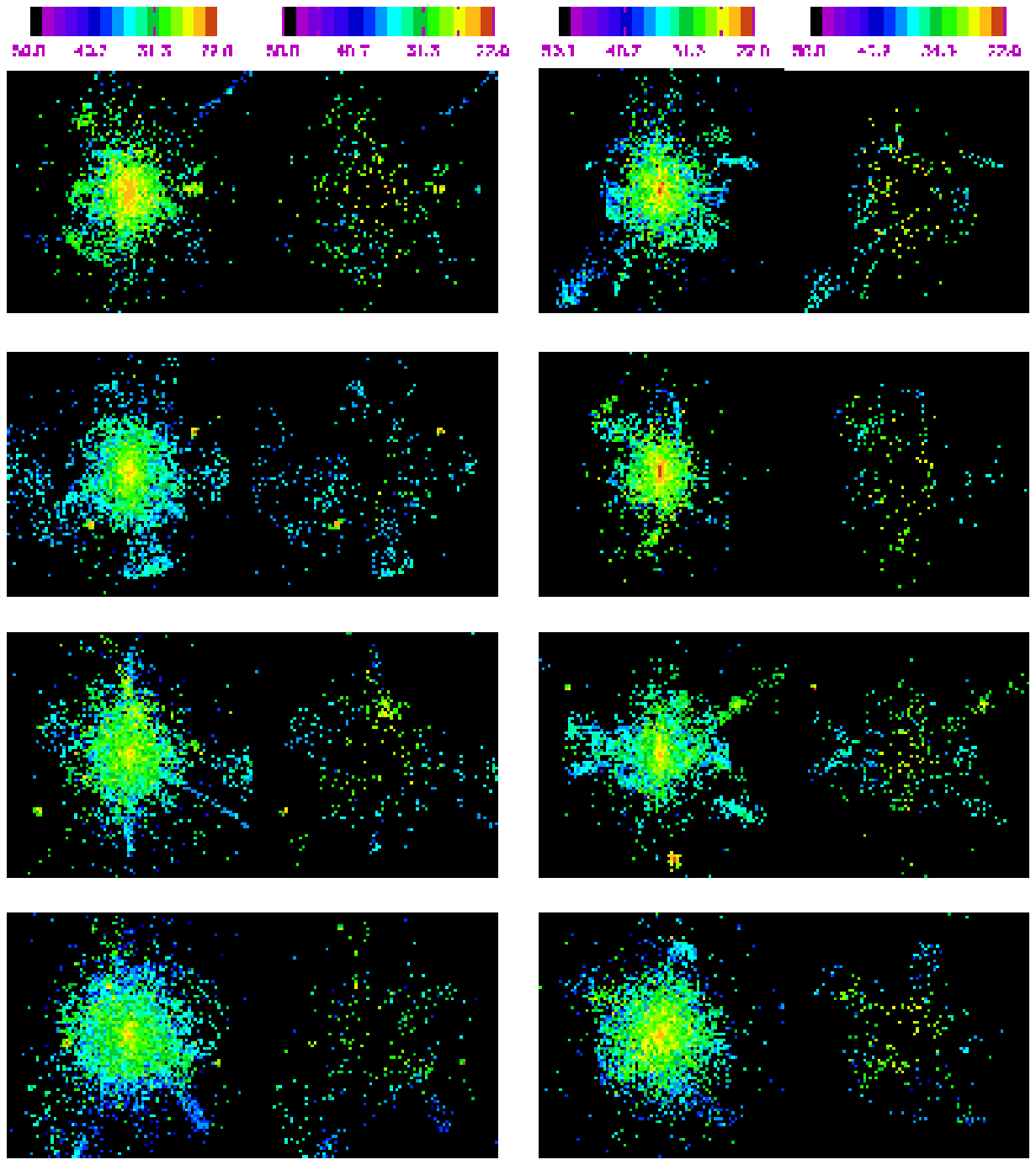}
\caption{\label{fig:metal_map_halos_weights}{V-band surface brightness maps for halos H1, H2, H3, H4, H5, H7, H8 and H10 (in order from top, left to right). Left sub-panels show the surface brightness map of all stars and right sub-panels the corresponding surface brightness of stars with \alphaFe $<0.05$. All maps span 300~kpc on the size.}}
\end{figure}

Figure  \ref{fig:average_afe_feh} plots  the  weight averaged \alphaFe
versus \FeH values for satellites  accreted onto halo H1, highlighting
the times of accretion. This shows that the lowest \alphaFe values are
associated with   recently accreted  satellites, these  systems having
more time available to   form stellar populations with \alphaFe  below
solar.     This   is      more     clearly    seen       in     Figure
\ref{fig:metal_en_angmom_chem_halo05}, where we plot the  distribution
of stars in the $E - L_{z}$ and $E - L$ planes, separated in different
\alphaFe  ranges. Most of  the stars with \alphaFe  $<0.05$ are in the
high  energy and    angular momentum regime,   associated  with recent
accretions.

This also suggests that by selecting an \alphaFe $<0.05$ cut, one can effectively remove the bulk of the old, well mixed halo and identify only the recently accreted streams. Figure \ref{fig:metal_map_halos_weights} illustrates this with a comparison between the surface brightness maps of some of the simulated halos, first by including all stars and alongside including only those stars with \alphaFe $<0.05$. With this cut, the central regions are effectively devoid of the well mixed streams and contain only the cold streams.

How feasible is our selection method for detecting streams in the halo? Figure \ref{fig:metal_map_halos_weights} shows that even without the \alphaFe cut the tidal streams are faint.  Only a few of these streams, typically 1 or 2 per galaxy, can be detected with current capabilities which have a limiting surface brightness in V-band of $\mu_{V} \simeq 30-31$ mag/arcsec$^{2}$.  The proposed method can be used to highlight a few cold streams of about this limiting magnitude in the inner halo. 

The prospects for detecting giant stars are also reasonable.  Assuming the luminosity of halo stars is similar to the one of globular cluster M12 \citep{hargis04}, a $10^{9} L_{\odot}$ halo will contain a few $10^{6}$ giants (out of these, $10^{5}$ giants will be 1 V-band mag above the horizontal branch (HB) or about $10^{4}$ giants 2~V-band mag above the HB, respectively). According to our models,  $1-10 \%$ of the stars meet the \alphaFe $<0.05$ cut. This implies, for example, that a halo survey of $10^{5}$ giants with 1mag above the HB will contain $\sim 10^{3} - 10^{4}$ giants in cold streams.

\section{Comparisons with Other Studies}
\label{sec:comparisons}

\subsection{Stellar Halo Metallicities}
\label{sec:feh}

A series of
recent observational studies have found evidence in support of the
idea that the metallicity/age of the Milky Way stellar 
halo is peculiar.
It seems that most other   bright spiral galaxies  are  more
metal rich than  the Milky  Way \citep[e.g.][]{mouhcine05a}.  
M31 is a  prime  example, but recent
observations show   that this is   also  the case  for  several spiral
galaxies outside  the Local  Group  \citep{mouhcine05b}, as  it is for
about  1000 spiral galaxies in  the SDSS sample,  as inferred from the
average  of their  colors   \citep{zibetti04}. Similarly, in terms  of
their stellar composition, the   observations suggest that  all  halos
have  an underlying, uniform population   of old metal  poor stars and
that most luminous  galaxies  have additional metal rich  populations,
which      can  be    either      old  or     of   intermediate    age
\citep[e.g.][]{harris99,harris02,bellazzini03,mouhcine05b,gallazzi05}. As
a bright galaxy but with  an old and metal  poor stellar halo, the
Milky Way may indeed be somewhat rare.
Indeed, none   of   our   11  simulated   stellar  halos   have
metallicities as low as that of the Milky Way
\avFeH$_{MW} \simeq -1.5$, although our most metal poor halo is close
\avFeH $\simeq -1.3$ (see   Figure  \ref{fig:metal_halo_averages}).
Our interpretation would be that the Milky Way is unusually lacking
in massive satellite mergers  at  early  times.  
While rare, this would   not   be unthinkable  within
the expectations of $\Lambda$CDM since the average number of accreted
massive satellites is small \citep{zb03}.

As shown in Figure \ref{fig:metal_halo_averages}, 
the metallicities of our simulated halos have a spread
of   less  than  1~dex.  This disagrees
somewhat with the results of \cite{renda05}, who 
report that their simulated halos  have a larger \FeH spread,  about 1.5~dex, when restricted to have
Milky Way-type luminosities.
More specifically, the \cite{renda05} simulations produce
 Milky Way-type halos  as
metal poor as \avFeH  $ \simeq -2$.
We suspect that the difference arises from
the fact that \cite{renda05} allow stellar halo stars to originate via
in situ star formation in their simulations, while our stellar halos
form entirely from accreted dwarfs.  Thus the metallicity spread in observed
halos may provide an interesting avenue for testing the idea of accreted vs.
in situ stellar halo formation.  
The small spread in stellar halo metallicities seen by 
\cite{mouhcine05b} may favor an accretion-dominated model, although the
evidence is weak as of now.
 A second difference may arise from
the cosmological conditions in \cite{renda05}, which are set via 
a semi-cosmological top-hat sphere with standard CDM fluctuations, while
ours rely on Extended Press Schechter within an $\Lambda$CDM framework.  
It is interesting to note that our restriction on accretion histories
to be those without recent major mergers would (if anything)
bias our distributions to more metal poor halos
 --- just the opposite bias
that would be required to match reconcile our results with those of
\cite{renda05}.

\subsection{Spatial Distributions}
\label{sec:spatial}

Our models predict that tidal  streams from massive satellites should
stand   out in  stellar overdensities   and   be more  metal rich. The
available observational evidence seems to  support our finding. A wide
survey of the  inner M31 halo reveals   an increase in  metallicity in
fields associated with   the giant stellar stream  \citep{ferguson02}.
Numerical  modeling of the  stream finds that the progenitor satellite
was  indeed    massive,   $M_{*}     \sim  10^{9}    \,     M_{\odot}$
\citep{font04,fardal05}.  Inhomogeneities  in color magnitude diagrams
are also seen  in the M31 halo and  are believed to be associated with
satellite accretions \citep{brown03,ferguson05}.

\begin{figure}
\figurenum{12}
\epsscale{2}
\plottwo{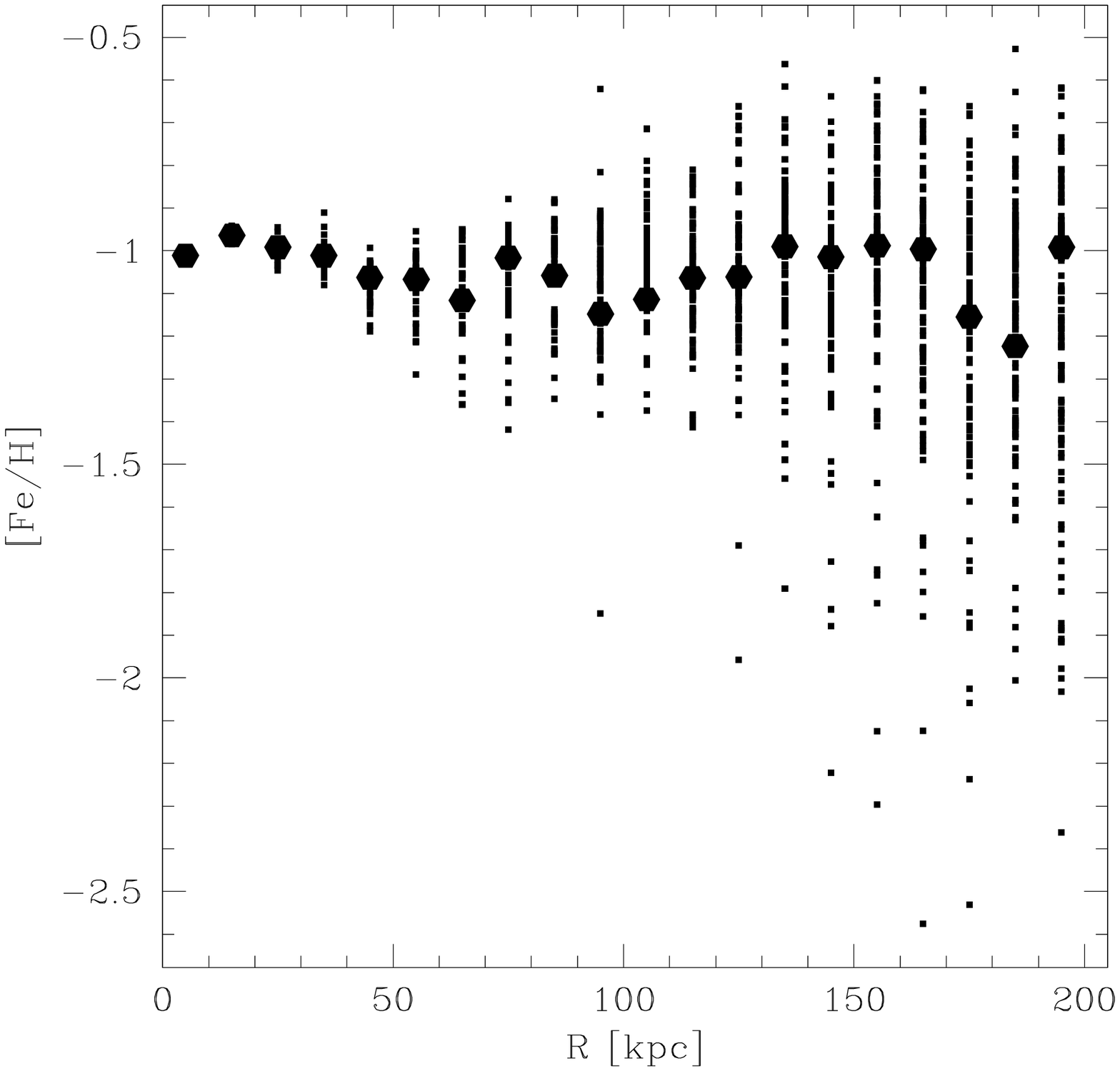}{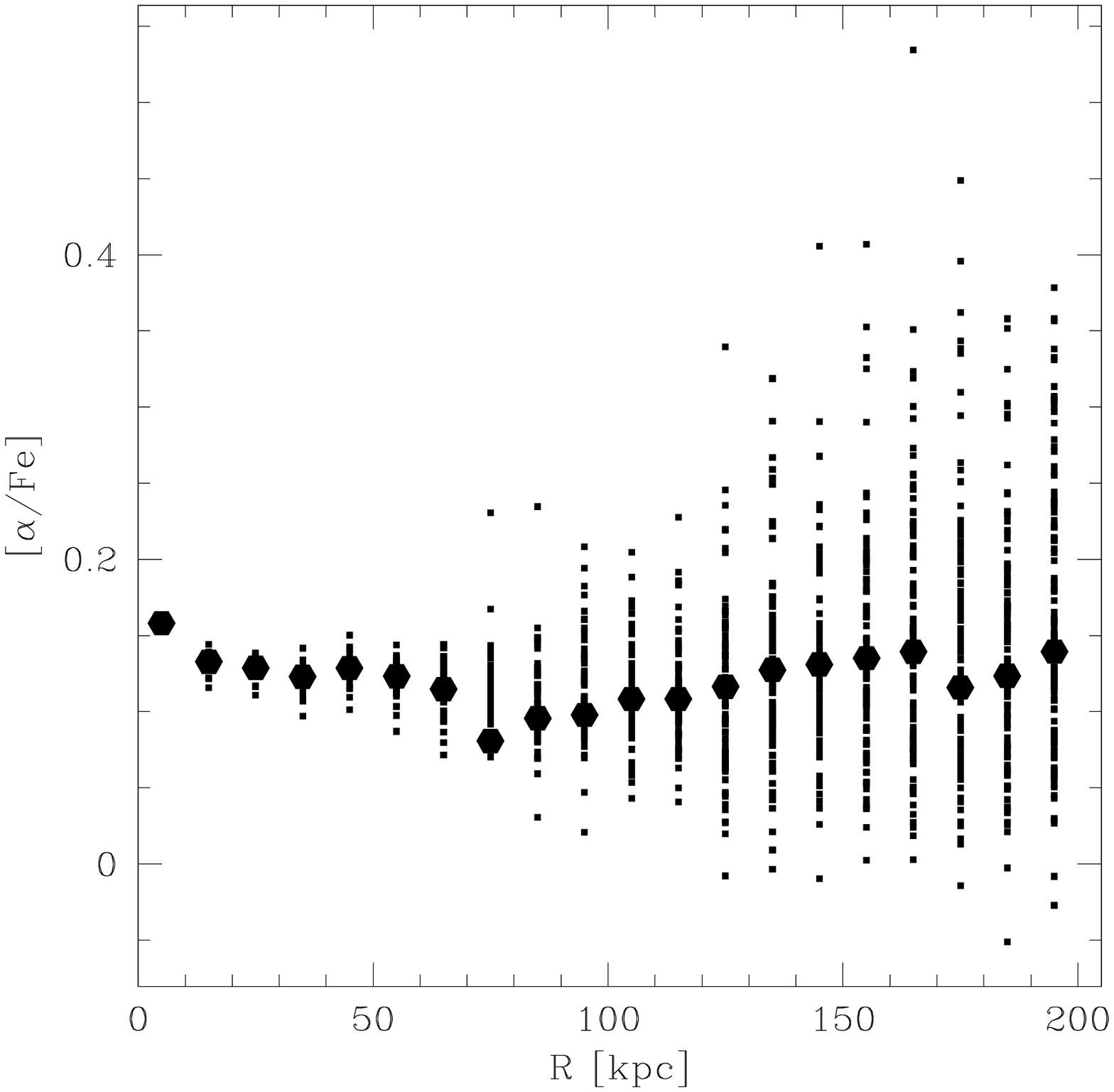}
\caption{\label{fig:metal_cyl_gradient} {\FeH and \alphaFe distributions versus cylindrical radius $R$ in Halo H1, ($R=\sqrt{x^{2}+z^{2}}$).  Large pentagons correspond to weight averaged \FeH or \alphaFe values in cylindrical shells of width $dR=10$~kpc.  Small squares correspond to weight averaged \FeH or \alphaFe values at equidistant azimuthal angles $\phi \in [0 , 2\pi]$ in a given cylindrical shell. (Note that the number of angles $\phi$ probed in each shell increases with $R,$ so as to preserve the size of the mock observational fields).
}}
\end{figure}

While    metallicity  inhomogeneities  are    commonly  detected,  the
observational evidence   regarding   radial  gradients    is so    far
inconclusive.        Studies        of          globular      clusters
\citep[e.g.][]{searle78,zinn93} or of the low metallicity stars in the
Galaxy \citep{beers05}  do    not show obvious    gradients in  \FeHt.
However, some  studies report  a decline of   \alphaFe ratios  of halo
stars with  radial distance in  the  Galaxy \citep{nissen97}.  In M31,
measurements in several  fields extending up to  $40$~kpc  in the halo
are consistent with a  small,  less than  1~dex, \FeH radial  gradient
\citep{bellazzini03}.

Our results suggest that metallicity gradients of less than 1~dex over
distances of  a few tens  of kpc are  consistent with the hierarchical
structure formation scenario. However,  for a more accurate comparison
between our models and observations, we have  to take into account the
methods of measurement used in observations.  For example, in external
galaxies  radial   distances cannot  be  currently  determined.   What
observations generally measure are the separations from the centers of
galaxies, which are projected radii on the plane of the sky. Moreover,
current  observations   probe small fields   in the  stellar  halos at
different projected radii \citep[e.g.][]{bellazzini03}. We  illustrate
this with a  mock  observation  of  one of   our  halos (H1).   Figure
\ref{fig:metal_cyl_gradient} shows the \FeH and \alphaFe distributions
versus  the  projected radius $R$    on the plane  $(x,z)$.  The large
symbols are averages in   cylindrical shells, while the  small symbols
show the decomposition of these values at equidistant azimuthal angles
in  each cylindrical  shell.   Thus each small  symbol  approximates a
projected field of observation  a few kpc  on the size, similar to the
size   of  observational fields.  Figure  \ref{fig:metal_cyl_gradient}
demonstrates that a large spread in either \FeH  or \alphaFe may exist
at a given $R$ and therefore a random sampling  with small fields over
different  projected radii  can   result in a   detection of  spurious
abundance gradients.  Our  results  therefore  caution that  a   wider
spatial sampling of halos is necessary in order to confirm the current
detections of abundance gradients.

\section{Conclusions}
\label{sec:conclusions}

The  motivation  of   this work was   to provide   predictions of  the
hierarchical formation paradigm that can be  tested with the data from
current and  upcoming astrometric and  spectroscopic observations.  In
this  study we have explored  the global properties as the phase-space
distribution of the \FeH and  \alphaFe chemical abundances in a series
of Milky  Way-type stellar  halos  formed in a  $\Lambda$CDM Universe.
Our results can be summarized as follows:
 
\vskip 4pt 
- For a fixed dark matter halo mass, we predict that stellar 
halo metallicities increase with stellar halo mass. The resultant
\FeH $- M_{*}$ relation for stellar halos arises because 
high-mass stellar halos tend to have accreted a larger number of
massive satellite progenitors.  The relation propagates
hierarchically because massive progenitors 
are more metal rich than their low-mass counterparts.

\vskip 4pt
- Chemical abundance distributions provide important links
to the minor  merger
histories of  galaxies.  For    example,  the observed  metallicity
differences between  the (metal poor) Milky Way  and (metal rich)
M31 stellar halos may arise because
M31 destroyed
one or two more massive satellites than the Milky Way.

\vskip 4pt 
- In agreement  with  observations, we  find that  all simulated halos
contain an underlying population of old metal poor stars.  More
massive halos contain additional intermediate age populations accreted
more recently.

\vskip 4pt
- In contrast with the Milky Way, some of  our halo models contain old
stellar populations which   are  metal rich,  originating   in massive
satellites ($M_{*}  \sim  10^{9} M_{\odot}$) accreted   early on.  The
metallicity constraints  seems  to suggest that the  Milky  Way has an
atypical   mass accretion  history, lacking in massive early
accretion events,   compared with  other  galaxies of
a similar mass.
 
\vskip 4pt 
- We predict metallicity gradients in \FeH  and \alphaFe in our
{\it hierarchically-formed}
stellar halos that are often non-negligible.
Gradients extend typically  over a few tens  of kpc and may be as large as
$0.5$~dex in \FeH and $0.2$~dex in \alphaFet.

\vskip 4pt 
- We predict that most metal poor stars in the Galactic halo are buried within
the central $\sim 5$kpc of the Galaxy.  This will likely be important
for efforts to use the observed count of low metallicity stars to constrain
models of cosmic reionization and the transition from Pop III to Pop II star
formation \citep[e.g.][]{tumlinson04}.

\vskip 4pt 
- We suggested a method to  identify cold tidal streams  in
stellar halos  by selecting stars   with \alphaFe $  \le 0.05$,  which
trace streams from recently accreted satellites. If used to complement
the common kinematical  identification methods, the proposed selection
criterion may improve the detection of tidal streams.

\vskip4pt
Combining kinematic and chemical abundance information  proves to be a
powerful  tool  for understanding how   galaxies  form and  evolve. In
anticipation   to the wealth  of data  to be  delivered by current and
future Galactic  surveys, it is  important  to develop and  improve on
numerical simulations of  the chemical  evolution of  the Galaxy in  a
realistic  cosmological context $-$  such as   those presented in  our
study $-$ and to devise criteria to identify more tidal streams.

While our methods successfully reproduce the gross chemical abundance properties of the Galactic halo and satellite dwarf galaxies, our treatments of dissipational physics including star formation, gas cooling, and energetic feedback mechanisms may be too simplistic to recover the detailed properties of the modeled systems with high accuracy.  For instance, the assumed truncation of star formation in dwarf
systems after they accrete into their host galaxy is very approximate. Gravitational torques applied by the central galaxy and other satellite systems may induce episodes of efficient star formation in the
dwarfs we model, especially near pericentric passage. Such bursts occurred in the LMC and SMC, as inferred from their stellar populations \citep[e.g.][]{smecker-hane02,harris04}, and should be modeled
if more accurate chemical enrichment histories for dwarf systems are to be achieved.

These complications speak to the need for high-resolution hydrodynamical simulations of the formation of the Galactic stellar halo and dwarf satellite population.  While recent hydrodynamical simulations have begun to address this issue \citep[e.g.][]{brook04},  they have not yet matched the mass resolution necessary to capture the entire population of dwarfs galaxies we follow in our collisionless simulations nor the full complexity of our chemical enrichment model.  We acknowledge the importance of the previous work and look forward to increasingly  sophisticated hydrodynamical simulations of the cosmological formation of the Galactic stellar halo.

\acknowledgements
The authors wish to thank Annette Ferguson, Tom Brown, Harry Ferguson, Raja Guhathakurta and Chris Purcell for helpful discussions.  A.F and K.V.J's contributions were supported through NASA grant NAG5-9064 and NSF CAREER award AST-0133617. JSB is supported by the Center for Cosmology at UC Irvine.

\bibliographystyle{apj}

\begin{thebibliography}{109}
\expandafter\ifx\csname natexlab\endcsname\relax\def\natexlab#1{#1}\fi

\bibitem[{{Abadi}, {Navarro} \& {Steinmetz}}(2005)]{abadi05}{Abadi}, M.~G.,  {Navarro}, J.~F., \& {Steinmetz}, M. submitted to \mnras \, (astro-ph/0506659)

\bibitem[{{Altmann}, {Catelan}, \& {Zoccali}(2005)}]{altmann05}{Altmann}, M., {Catelan}, M. \& {Zoccali}, M. 2005 \aap, 439, L5

\bibitem[{{Beers} et~al.(2005)}]{beers05} {Beers}, T.~C. et~al., to appear in ``From Lithium to Uranium: Elemental Tracers of Early Cosmic Evolution", IAU Symposium 228, V. Hill, P. Francois \& F. Primas, eds. (astro-ph/0508423)

\bibitem[{{Bekki} \& {Chiba}(2001)}]{bekki01} {Bekki}, K. \& {Chiba}, M. 2001, \apj, 558, 666
 
\bibitem[{{Bellazzini} et~al.(2003)}]{bellazzini03} {Bellazzini}, M., {Cacciari}, C., {Federici}, L., {Fusi Pecci}, F., {Rich}, M. 2003 \aap, 405, 867

\bibitem[{{Bland-Hawthorn} \& {Freeman}(2000)}]{bland-hawthorn00} {Bland-Hawthorn}, J. \& {Freeman}, K. 2000, Science, 287, 79

\bibitem[{{Blumenthal} et~al.(1984)}]{blumenthal84} {Blumenthal}, G.~R., {Faber}, S.~M., {Primack}, J.~R. \& {Rees}, M.~J. 1984, Nature, 311, 517

\bibitem[{{Brodie} \& {Huchra}(1991)}]{brodie91} {Brodie}, J. \& {Huchra}, J.~P. 1991, \apj, 379, 157

\bibitem[{{Brook} et~al.(2003)}]{brook03} {Brook}, C.~B., {Kawata}, D., {Gibson}, B.~K., \& {Flynn}, C. 2003, \apj, 585, L125

\bibitem[{{Brook} et~al.(2004)}]{brook04} {Brook}, C.~B., {Kawata}, D., {Gibson}, B.~K., \& {Flynn}, C. 2004, \apj, 349, 52

\bibitem[{{Brown} et~al.(2003)}]{brown03} {Brown}, T. ~M et~al. 2003, \apj, 592, 17L

\bibitem[{{Bullock} \& {Johnston}(2005)}]{bullock05}  {Bullock}, J.~S. \& {Johnston}, K.~V., \apj, accepted (astro-ph/0506467)

\bibitem[{{Bullock}, {Kravtsov}, \& {Weinberg}(2001)}]{bullock01} {Bullock}, J.~S., {Kravtsov}, A.~V., \& {Weinberg}, D.~H. 2001, \apj, 548, 33

\bibitem[{{Carney} et~al.(1996)}]{carney96} {Carney}, B.~W., {Laird}, J.~B., {Latham}, D.~W., {Aguilar}, L.~A.  1996, \aj, 112, 668

\bibitem[{{Chiba} \& {Beers}(2000)}]{chiba00} {Chiba}, M. \& {Beers}, T.~C. 2000, \aj, 119, 2843

\bibitem[{{Dekel} \& {Silk}(1986)}]{dekel86} {Dekel}, A. \& {Silk}, J. 1986, \apj, 303, 39

\bibitem[{{Dekel} \& {Woo}(2003)}]{dekel03} {Dekel}, A. \& {Woo}, J. 2003, \mnras, 344, 1131

\bibitem[{Diemand}, {Madau} \& {Moore}(2005)]{diemand05} {Diemand}, J., {Madau}, P., \& {Moore}, B. 2005, \mnras, in press

\bibitem[{{Dinescu}(2002)}]{dinescu02} {Dinescu}, D.~I. 2002, ASPC, 265, 365

\bibitem[{{Eggen}, {Lynden-Bell} \& {Sandage}(1962)}]{eggen62} {Eggen}, O.~J., {Lynden-Bell}, D. \& {Sandage}, A.~R. 1962, \apj, 136, 748

\bibitem[{{Fardal} et~al.(2005)}]{fardal05} {Fardal}, M.~A., {Babul}, A., {Geehan}, J.~J., \& {Guhathakurta}, P., submitted to \mnras \, (astro-ph/0501241)

\bibitem[{{Ferguson} et~al.(2002)}]{ferguson02} {Ferguson}, A.~M.~N., {Irwin}, M.~J., {Ibata}, R.~A., {Lewis}, G.~F. \& {Tanvir}, N.~R. 2002, \aj, 124, 1452

\bibitem[{{Ferguson} et~al.(2005)}]{ferguson05} {Ferguson}, A.~M.~N. et~al. 2005, ApJL, 622,  109

\bibitem[{{Font} et~al.(2004)}]{font04} {Font}, A.~S., {Johnston}, K.~V., {Guhathakurta}, P., {Majewski}, S.~R., {Rich}, R.~M., \aj, accepted (astro-ph/0406146)

\bibitem[{{Font} et~al.(2005)}]{font05} {Font}, A.~S., {Johnston}, K.~V., {Bullock}, J.~S., \& {Robertson}, B.~E.,  \apj,  accepted (astro-ph/0507114)

\bibitem[{{Freeman} \& {Bland-Hawthorn}(2002)}]{freeman02} {Freeman}, K. \& {Bland-Hawthorn}, J. 2002, ARA\&A, 40, 487

\bibitem[{{Gallazzi} et~al.(2005)}]{gallazzi05} {Gallazzi}, A. et~al. 2005, \mnras, 362, 41

\bibitem[{{Gilmore} \& {Wyse}(1998)}]{gilmore98} {Gilmore}, G. \& {Wyse}, R.~F.~G. 1998, \aj, 116, 748 

\bibitem[{{Grebel} et~al.(2003)}]{grebel03} {Grebel}, E.~K., {Gallagher}, J.~S. \& {Harbeck}, D. 2003, \aj, 125, 1926

\bibitem[{{Guhathakurta} et~al.(2005a)}]{guhathakurta05a} {Guhathakurta}, P. et~al. 2005, submitted to Nature (astro-ph/0502366)

\bibitem[{{Guhathakurta} et~al.(2005b)}]{guhathakurta05b} {Guhathakurta}, P. et~al., \aj \, accepted (astro-ph/0406145)

\bibitem[{{Hargis}, {Sandquist} \& {Bolte}(2004)}]{hargis04}{Hargis}, J.~R., {Sandquist}, E~L., {Bolte}, M. 2004, \apj, 608, 243

\bibitem[{{Harris} et~al.(1999)}]{harris99} {Harris}, G.~L.~H., {Harris}, W.~E. \& {Poole}, G.~B. 1999, \aj, 117, 855

\bibitem[{{Harris} \& {Harris}(2002)}]{harris02} {Harris}, W.~E. \& {Harris}, G.~L.~H. 2002, \aj, 123, 3108

\bibitem[{{Harris} \& {Zaritsky}(2004)}]{harris04}{Harris}, J. \& {Zaritsky}, D. 2004, \aj, 127, 1531

\bibitem[{{Helmi} \& {de Zeeuw}(2000)}]{helmi00} {Helmi}, A. \& {de Zeeuw}, P.~T. 2000, \mnras, 319, 657

\bibitem[{{Helmi} et~al.(2005)}]{helmi05} {Helmi}, A., {Navarro}, J.~F., {Nordstr{\"o}m}, B., {Holmberg}, J., {Abadi}, M.~G. \& {Steinmetz}, M. , \mnras, in press (astro-ph/0505401)

\bibitem[{{Helmi} et~al.(1999)}]{helmi99} {Helmi}, A., {White}, S.~D.~M., {de Zeeuw}, P.~T., \& {Zhao}, H. 
1999, Nature, 402, 53

\bibitem[{Helmi}, {White} \& {Springel}(2003)]{helmi03} {Helmi}, A., {White}, S.~D.~M., \& {Springel}, V. 2003, \mnras, 339, 834

\bibitem[{{Ibata} et~al.(2001a)}]{ibata01a} {Ibata}, R.~A., {Irwin}, M.~J., {Ferguson}, A.~M.~N., {Lewis}, G.~F., \& {Tanvir}, N. 2001, Nature, 412, 49 (a)

\bibitem[{{Ibata} et~al.(2004)}]{ibata04} {Ibata}, R.~A., {Chapman}, S., {Ferguson}, A.~M.~N., {Irwin}, M., {Lewis}, G.~F., \& {McConnachie}, A.. 2004, \mnras, 351, 117

\bibitem[{{Ibata} et~al.(2005)}]{ibata05} {Ibata}, R., {Chapman}, S., {Ferguson}, A.~M.~N. , {Lewis}, G., {Irwin}, M., {Tanvir}, N, 2005, submitted to \apj \, (astro-ph/0504164)

\bibitem[{{Irwin} et~al.(2005)}]{irwin05} {Irwin}, M., {Ferguson}, A.~M.~N., {Ibata}, R., {Lewis}, G, {Tanvir}, N. 2005, \apj, 628, L105

\bibitem[{Johnston}(1998)]{johnston98} {Johnston}, K.~V. 1998, \apj, 495, 297

\bibitem[{{Kalirai} et~al.(2005)}]{kalirai05} {Kalirai}, J~S. et al. 2005, \apj, accepted (astro-ph/0512161).

\bibitem[{{Kauffmann} et~al.(1999)}]{kauffmann99} {Kauffmann}, G., {Colberg}, J.~M., {Diaferio}, A., \& {White}, S.~D.~M. 1999, \mnras, 303, 188

\bibitem[{{Laird} et~al.(1988)}]{laird88} {Laird}, J.~B., {Carney}, B.~W., {Rupen}, M.~P., \& {Latham}, D.~W. 1988, \aj, 96, 1908

\bibitem[{{Larson}(1974)}]{larson74} {Larson}, R.~B. 1974, \mnras, 169, 229

\bibitem[{{Majewski}, {Munn} \& {Hawley}(1996)}]{majewski96} {Majewski}, S.~R., {Munn}, J.~A. \& {Hawley}, S.~L. 1996, \apj, 459, L73

\bibitem[{{Mayer} et~al.(2002)}]{mayer02} {Mayer}, L., {Moore}, B., {Quinn}, T., {Governato}, F., {Stadel}, J. 2002, \mnras, 336, 119

\bibitem[{{Moore} et~al.(2005)}]{moore05} {Moore}, B., {Diemand}, J., {Madau}, P., {Zemp}, M., \& {Stadel}, J., submitted to \mnras \, (astro-ph/0510370)

\bibitem[{{Mouhcine} et~al.(2005a)}]{mouhcine05a} {Mouhcine}, M., {Ferguson}, H.~C., {Rich}, R.~M., {Brown}, T.~M., \& {Smith}, T.~E.  2005a, \apj, 633, 821

\bibitem[{{Mouhcine} et~al.(2005b)}]{mouhcine05b} {Mouhcine}, M., {Ferguson}, H.~C., {Rich}, R.~M., {Brown}, T.~M., \& {Smith}, T.~E.  2005b, \apj, 633, 828

\bibitem[{{Navarro} et~al.(2004)}]{navarro04} {Navarro}, J.~F., {Helmi}, A. \& {Freeman}, K.~C. 2004, \apj, 601, L43

\bibitem[{{Nissen} \& {Schuster}(1991)}]{nissen91} {Nissen}, P.~E. \& {Schuster}, W.~J. 1991, \aap, 251, 457

\bibitem[{{Nissen} \& {Schuster}(1997)}]{nissen97} {Nissen}, P.~E. \& {Schuster}, W.~J. 1997, \aap, 326,751

\bibitem[{{Perryman} et~al.(2001)}]{perryman01} {Perryman}, M.~A.~C. et~al. 2001, \aap, 369, 339

\bibitem[{{Reitzel} \& {Guhathakurta}(2002)}]{reitzel02}{Reitzel}, D.~B. \& {Guhathakurta}, P. 2002, \aj, 124, 234

\bibitem[{{Renda} et~al.(2005)}]{renda05} {Renda}, A. et~al. 2005, \mnras, 363, L16

\bibitem[{{Robertson} et~al.(2005)}]{robertson05} {Robertson}, B., {Bullock}, J.~S., {Font}, A.~S., {Johnston}, K.~V. \&  {Hernquist}, L. 2005, \apj, 632, 872

\bibitem[{{Searle}(1977)}]{searle77} {Searle}, L. 1977, in ``The Evolution of Galaxies and Stellar Populations'', ed. B. M. Tinsley, R. B. Larson, p. 219, New Haven: Yale Univ. Press

\bibitem[{{Searle} \& {Zinn}(1978)}]{searle78} {Searle}, L. \& {Zinn}, R. 1978, \apj, 225, 357

\bibitem[{{Smecker-Hane} et~al.(2002)}]{smecker-hane02}{Smecker-Hane}, T.~A., {Cole}, A.~A., {Gallagher}, J.~S. \& {Stetson}, P.~B. 2002, \apj, 566, 239

\bibitem[{{Springel} et~al.(2001)}]{springel01} {Springel}, V., {White}, S.~D.~M., {Tormen}, G., {Kauffmann}, G. 2001, \mnras, 328, 726

\bibitem[{{Steinmetz}(2003)}]{steinmetz03} {Steinmetz}, M., 2003, in ``GAIA Spectroscopy: Science and Technology", ASP Conference Proceedings, edited by Ulisse Munari, Vol. 298, p.381

\bibitem[{{Tinsley}(1980)}]{tinsley80} {Tinsley}, B.~M. 1980, FCPh, 5, 287

\bibitem[{{Tremonti} et~al.(2004)}]{tremonti04} {Tremonti}, C.~A. et~al. 2004, \apj, 613, 898

\bibitem[Tumlinson et al.(2004)]{tumlinson04} {Tumlinson}, J., 
{Venkatesan}, A., \& {Shull}, J.~M.\ 2004, \apj, 612, 602 

\bibitem[{{White} \& {Rees}(1978)}]{white78} {White}, S.~D.~M. \& {Rees}, M. 1978, \mnras, 183, 341

\bibitem[{{Wyse}(2001)}]{wyse01} {Wyse}, R.~F.~G., 2001, in ``Galactic disks and disk galaxies'' (ed. J. Funes \& E. Corsini) ASP Conference Series vol 230, (ASP, San Francisco) 71

\bibitem[{{Zaritsky}, {Kennicutt} \& {Huchra}(1994)}]{zaritsky94} {Zaritsky}, D., {Kennicutt}, R.~C., {Huchra}, J.~P. 1994, \apj, 420, 87

\bibitem[Zentner \& Bullock(2003)]{zb03} {Zentner}, A.~R., \& 
{Bullock}, J.~S.\ 2003, \apj, 598, 49 
 
\bibitem[{{Zibetti}, {White} \& {Brinkmann}(2004)}]{zibetti04} {Zibetti}, S., {White}, S.~D.~M., \& {Brinkmann}, J. 2004, \mnras, 347, 556

\bibitem[{{Zinn}(1993)}]{zinn93} {Zinn}, R. 1993, in ``The globular clusters-galaxy connection''. Astronomical Society of the Pacific (ASP) Conference Series, edited by Graeme H. Smith, and Jean P. Brodie, Vol. 48, p.38

\end{thebibliography}

\end{document}